\documentclass[twocolumn]{aastex63}

\accepted{March 4, 2021}
\submitjournal{ApJ}

\usepackage[T1]{fontenc}
\usepackage[utf8]{inputenc}
\usepackage{amsmath}
\usepackage{lmodern}
\usepackage{numname}
\usepackage{xspace}

\shorttitle{Twins Embedding II}
\shortauthors{Boone et al.}
\graphicspath{{./}{figures/}}

\makeatletter
\def\input@path{{./}{latex/}}
\makeatother

\defcitealias{fakhouri15}{F15}
\defcitealias{branch06}{B06}
\defcitealias{rigault18}{R18}

\newcommand{\Usnf}{$U_\mathrm{SNf}$\xspace}
\newcommand{\Bsnf}{$B_\mathrm{SNf}$\xspace}
\newcommand{\Vsnf}{$V_\mathrm{SNf}$\xspace}
\newcommand{\Rsnf}{$R_\mathrm{SNf}$\xspace}
\newcommand{\Isnf}{$I_\mathrm{SNf}$\xspace}

\newcommand{\rawrbtlmagstd}{0.131 $\pm$ 0.010}
\newcommand{\rawrbtlmagnmad}{0.108 $\pm$ 0.013}
\newcommand{\rawrbtlmagstdsaltcuts}{0.131 $\pm$ 0.011}

\newcommand{\rbtlgprms}{0.101 $\pm$ 0.007}
\newcommand{\rbtlgpnmad}{0.083 $\pm$ 0.010}
\newcommand{\rbtlgpintdisp}{0.073 $\pm$ 0.008}
\newcommand{\rbtlgpkernelamp}{0.164 $\pm$ 0.082}
\newcommand{\rbtlgpkernellengthscale}{5.54 $\pm$ 3.31}

\newcommand{\rbtlgprv}{2.40 $\pm$ 0.16}

\newcommand{\rbtlgprmssaltcut}{0.100 $\pm$ 0.008}

\newcommand{\saltgprms}{0.118 $\pm$ 0.008}
\newcommand{\saltgpnmad}{0.105 $\pm$ 0.013}
\newcommand{\saltgpintdisp}{0.085 $\pm$ 0.010}
\newcommand{\saltgpkernelamp}{0.380 $\pm$ 0.230}
\newcommand{\saltgpkernellengthscale}{8.63 $\pm$ 5.33}

\newcommand{\saltgpcolor}{2.81 $\pm$ 0.15}

\newcommand{\pecvelcontribution}{0.055}
\newcommand{\rbtlgppvremrms}{0.084 $\pm$ 0.009}

\newcommand{\rbtlgpcomppvremrms}{0.084 $\pm$ 0.009}
\newcommand{\saltcomppvremrms}{0.129 $\pm$ 0.014}

\newcommand{\nummanifoldsne}{173}

\newcommand{\numsnftraincombined}{97}
\newcommand{\numsnfvalid}{76}

\newcommand{\numsnredshift}{5}
\newcommand{\numlowredshift}{24}
\newcommand{\numhighav}{17}

\newcommand{\nummagsne}{134}

\newcommand{\numbadsalt}{18}
\newcommand{\numsaltsne}{155}

\newcommand{\saltparamalpha}{0.148 $\pm$ 0.011}
\newcommand{\saltparambeta}{2.71 $\pm$ 0.16}
\newcommand{\saltparamsigmaint}{0.118 $\pm$ 0.016}
\newcommand{\saltparamrms}{0.140 $\pm$ 0.012}
\newcommand{\saltparamwrms}{0.140 $\pm$ 0.013}
\newcommand{\saltparamnmad}{0.106 $\pm$ 0.013}

\newcommand{\saltparammindisp}{0.132}
\newcommand{\saltparammaxdisp}{0.173}

\newcommand{\saltfirstcompdiff}{0.229 $\pm$ 0.045}

\newcommand{\hoststepfitrbtllssfr}{0.066 $\pm$ 0.022}
\newcommand{\hoststepfitsaltlssfr}{0.121 $\pm$ 0.027}
\newcommand{\hoststepgmmrbtllssfr}{0.047 $\pm$ 0.018}
\newcommand{\hoststepgmmsaltlssfr}{0.093 $\pm$ 0.022}

\newcommand{\hoststeppeculiarsaltlssfr}{0.101 $\pm$ 0.031}
\newcommand{\hoststepfitrbtlmass}{0.040 $\pm$ 0.020}
\newcommand{\hoststepfitsaltmass}{0.092 $\pm$ 0.024}
\newcommand{\hoststepgmmrbtlmass}{0.032 $\pm$ 0.018}
\newcommand{\hoststepgmmsaltmass}{0.082 $\pm$ 0.021}

\newcommand{\hoststeppeculiarsaltmass}{0.059 $\pm$ 0.027}

\newcommand{\hoststepdifflssfrsignificance}{3.7}

\newcommand{\saltcompnamea}{SN2013be}
\newcommand{\saltcompnameb}{PTF11mkx}
\newcommand{\saltcompxonea}{$0.443 \pm 0.171$}
\newcommand{\saltcompxoneb}{$0.558 \pm 0.132$}
\newcommand{\saltcompca}{$0.025 \pm 0.028$}
\newcommand{\saltcompcb}{$0.042 \pm 0.027$}
\newcommand{\saltcompcoorda}{$-0.26$}
\newcommand{\saltcompcoordb}{$+5.17$}

\newcommand{\saltcompmagdiff}{$0.311 \pm 0.096$}
\newcommand{\saltcompmanifoldmagdiff}{$-0.002 \pm 0.104$}

\newcommand{\saltrbtlrmsdiff}{0.040 $\pm$ 0.011}
\newcommand{\saltrbtlrmsdiffsig}{3.7}

\begin{document}

\title{The Twins Embedding of Type Ia Supernovae II: Improving Cosmological Distance Estimates}

\correspondingauthor{Kyle Boone}
\email{kyboone@uw.edu}

\author[0000-0002-5828-6211]{K.~Boone}
\affiliation{Physics Division, Lawrence Berkeley National Laboratory, 1 Cyclotron Road, Berkeley, CA, 94720, USA}
\affiliation{Department of Physics, University of California Berkeley, 366 LeConte Hall MC 7300, Berkeley, CA, 94720-7300, USA}
\affiliation{DIRAC Institute, Department of Astronomy, University of Washington, 3910 15th Ave NE, Seattle, WA, 98195, USA}

\author{G.~Aldering}
\affiliation{Physics Division, Lawrence Berkeley National Laboratory, 1 Cyclotron Road, Berkeley, CA, 94720, USA}

\author[0000-0002-0389-5706]{P.~Antilogus}
\affiliation{Laboratoire de Physique Nucl\'eaire et des Hautes Energies, CNRS/IN2P3, Sorbonne Universit\'e, Universit\'e de Paris, 4 place Jussieu, 75005 Paris, France}

\author[0000-0002-9502-0965]{C.~Aragon}
\affiliation{Physics Division, Lawrence Berkeley National Laboratory, 1 Cyclotron Road, Berkeley, CA, 94720, USA}
\affiliation{College of Engineering, University of Washington 371 Loew Hall, Seattle, WA, 98195, USA}

\author{S.~Bailey}
\affiliation{Physics Division, Lawrence Berkeley National Laboratory, 1 Cyclotron Road, Berkeley, CA, 94720, USA}

\author[0000-0003-0424-8719]{C.~Baltay}
\affiliation{Department of Physics, Yale University, New Haven, CT, 06250-8121, USA}

\author{S.~Bongard}
\affiliation{Laboratoire de Physique Nucl\'eaire et des Hautes Energies, CNRS/IN2P3, Sorbonne Universit\'e, Universit\'e de Paris, 4 place Jussieu, 75005 Paris, France}

\author[0000-0002-3780-7516]{C.~Buton}
\affiliation{Univ Lyon, Univ Claude Bernard Lyon~1, CNRS, IP2I~Lyon / IN2P3, UMR~5822, F-69622, Villeurbanne, France}
    
\author[0000-0002-5317-7518]{Y.~Copin}
\affiliation{Univ Lyon, Univ Claude Bernard Lyon~1, CNRS, IP2I~Lyon / IN2P3, UMR~5822, F-69622, Villeurbanne, France}

\author[0000-0003-1861-0870]{S.~Dixon}
\affiliation{Physics Division, Lawrence Berkeley National Laboratory, 1 Cyclotron Road, Berkeley, CA, 94720, USA}
\affiliation{Department of Physics, University of California Berkeley, 366 LeConte Hall MC 7300, Berkeley, CA, 94720-7300, USA}

\author[0000-0002-7496-3796]{D.~Fouchez}
\affiliation{Aix Marseille Univ, CNRS/IN2P3, CPPM, Marseille, France}

\author[0000-0001-6728-1423]{E.~Gangler}  
\affiliation{Univ Lyon, Univ Claude Bernard Lyon~1, CNRS, IP2I~Lyon / IN2P3, UMR~5822, F-69622, Villeurbanne, France}
\affiliation{Universit\'e Clermont Auvergne, CNRS/IN2P3, Laboratoire de Physique de Clermont, F-63000 Clermont-Ferrand, France}

\author[0000-0003-1820-4696]{R.~Gupta}
\affiliation{Physics Division, Lawrence Berkeley National Laboratory, 1 Cyclotron Road, Berkeley, CA, 94720, USA}

\author[0000-0001-9200-8699]{B.~Hayden}
\affiliation{Physics Division, Lawrence Berkeley National Laboratory, 1 Cyclotron Road, Berkeley, CA, 94720, USA}
\affiliation{Space Telescope Science Institute, 3700 San Martin Drive Baltimore, MD, 21218, USA}

\author{W.~Hillebrandt}
\affiliation{Max-Planck-Institut f\"ur Astrophysik,  Karl-Schwarzschild-Str. 1, D-85748 Garching, Germany}

\author[0000-0001-6315-8743]{A.~G.~Kim}
\affiliation{Physics Division, Lawrence Berkeley National Laboratory, 1 Cyclotron Road, Berkeley, CA, 94720, USA}

\author[0000-0001-8594-8666]{M.~Kowalski}
\affiliation{Institut f\"ur Physik,  Humboldt-Universitat zu Berlin, Newtonstr. 15, 12489 Berlin, Germany}
\affiliation {DESY, D-15735 Zeuthen, Germany}

\author{D.~K\"usters}
\affiliation {Department of Physics, University of California Berkeley, 366 LeConte Hall MC 7300, Berkeley, CA, 94720-7300, USA}
\affiliation {DESY, D-15735 Zeuthen, Germany}

\author[0000-0002-8357-3984]{P.-F.~L\'eget}
\affiliation{Laboratoire de Physique Nucl\'eaire et des Hautes Energies, CNRS/IN2P3, Sorbonne Universit\'e, Universit\'e de Paris, 4 place Jussieu, 75005 Paris, France}

\author{F.~Mondon}  
\affiliation{Universit\'e Clermont Auvergne, CNRS/IN2P3, Laboratoire de Physique de Clermont, F-63000 Clermont-Ferrand, France}

\author[0000-0001-8342-6274]{J.~Nordin}
\affiliation{Physics Division, Lawrence Berkeley National Laboratory, 1 Cyclotron Road, Berkeley, CA, 94720, USA}
\affiliation{Institut f\"ur Physik,  Humboldt-Universitat zu Berlin, Newtonstr. 15, 12489 Berlin, Germany}

\author[0000-0003-4016-6067]{R.~Pain}
\affiliation{Laboratoire de Physique Nucl\'eaire et des Hautes Energies, CNRS/IN2P3, Sorbonne Universit\'e, Universit\'e de Paris, 4 place Jussieu, 75005 Paris, France}

\author{E.~Pecontal}
\affiliation{Centre de Recherche Astronomique de Lyon, Universit\'e Lyon 1, 9 Avenue Charles Andr\'e, 69561 Saint Genis Laval Cedex, France}

\author{R.~Pereira}
\affiliation{Univ Lyon, Univ Claude Bernard Lyon~1, CNRS, IP2I~Lyon / IN2P3, UMR~5822, F-69622, Villeurbanne, France}

\author[0000-0002-4436-4661]{S.~Perlmutter}
\affiliation{Physics Division, Lawrence Berkeley National Laboratory, 1 Cyclotron Road, Berkeley, CA, 94720, USA}
\affiliation{Department of Physics, University of California Berkeley, 366 LeConte Hall MC 7300, Berkeley, CA, 94720-7300, USA}

\author[0000-0002-8207-3304]{K.~A.~Ponder}
\affiliation{Department of Physics, University of California Berkeley, 366 LeConte Hall MC 7300, Berkeley, CA, 94720-7300, USA}

\author{D.~Rabinowitz}
\affiliation{Department of Physics, Yale University, New Haven, CT, 06250-8121, USA}

 \author[0000-0002-8121-2560]{M.~Rigault} 
\affiliation{Institut f\"ur Physik,  Humboldt-Universitat zu Berlin, Newtonstr. 15, 12489 Berlin, Germany}
\affiliation{Univ Lyon, Univ Claude Bernard Lyon~1, CNRS, IP2I~Lyon / IN2P3, UMR~5822, F-69622, Villeurbanne, France}

\author[0000-0001-5402-4647]{D.~Rubin}
\affiliation{Physics Division, Lawrence Berkeley National Laboratory, 1 Cyclotron Road, Berkeley, CA, 94720, USA}
\affiliation{Department of Physics, University of Hawaii, 2505 Correa Rd, Honolulu, HI, 96822, USA}

\author{K.~Runge}
\affiliation{Physics Division, Lawrence Berkeley National Laboratory, 1 Cyclotron Road, Berkeley, CA, 94720, USA}

\author[0000-0002-4094-2102]{C.~Saunders}
\affiliation{Physics Division, Lawrence Berkeley National Laboratory, 1 Cyclotron Road, Berkeley, CA, 94720, USA}
\affiliation{Department of Physics, University of California Berkeley, 366 LeConte Hall MC 7300, Berkeley, CA, 94720-7300, USA}
\affiliation{Princeton University, Department of Astrophysics, 4 Ivy Lane, Princeton, NJ, 08544, USA}
\affiliation{Sorbonne Universit\'es, Institut Lagrange de Paris (ILP), 98 bis Boulevard Arago, 75014 Paris, France}

\author[0000-0002-9093-8849]{G.~Smadja}
\affiliation{Univ Lyon, Univ Claude Bernard Lyon~1, CNRS, IP2I~Lyon / IN2P3, UMR~5822, F-69622, Villeurbanne, France}

\author{N.~Suzuki}
\affiliation{Physics Division, Lawrence Berkeley National Laboratory, 1 Cyclotron Road, Berkeley, CA, 94720, USA}
\affiliation{Kavli Institute for the Physics and Mathematics of the Universe (WPI), The University of Tokyo Institutes for Advanced Study, The University of Tokyo, 5-1-5 Kashiwanoha, Kashiwa, Chiba 277-8583, Japan}

\author{C.~Tao}
\affiliation{Tsinghua Center for Astrophysics, Tsinghua University, Beijing 100084, China}
\affiliation{Aix Marseille Univ, CNRS/IN2P3, CPPM, Marseille, France}

\author[0000-0002-4265-1958]{S.~Taubenberger}
\affiliation{Max-Planck-Institut f\"ur Astrophysik, Karl-Schwarzschild-Str. 1, D-85748 Garching, Germany}

\author{R.~C.~Thomas}
\affiliation{Physics Division, Lawrence Berkeley National Laboratory, 1 Cyclotron Road, Berkeley, CA, 94720, USA}
\affiliation{Computational Cosmology Center, Computational Research Division, Lawrence Berkeley National Laboratory, 1 Cyclotron Road MS 50B-4206, Berkeley, CA, 94720, USA}

\author{M.~Vincenzi}
\affiliation{Physics Division, Lawrence Berkeley National Laboratory, 1 Cyclotron Road, Berkeley, CA, 94720, USA}
\affiliation{Institute of Cosmology and Gravitation, University of Portsmouth, Portsmouth, PO1 3FX, UK}

\collaboration{50}{The Nearby Supernova Factory}



\begin{abstract}

We show how spectra of Type~Ia supernovae (SNe~Ia) at maximum light can be used to improve
cosmological distance estimates. In a companion article, we used manifold learning
to build a three-dimensional parametrization of the intrinsic diversity of SNe~Ia at maximum light
that we call the ``Twins Embedding''. In this article, we
discuss how the Twins Embedding can be used to improve the standardization of
SNe~Ia. With a single spectrophotometrically-calibrated spectrum near maximum light, we can standardize
our sample of SNe~Ia with an RMS of \rbtlgprms~mag, which corresponds to \rbtlgppvremrms~mag
if peculiar velocity contributions are removed and \rbtlgpintdisp~mag if a larger reference sample were obtained.
Our techniques can standardize the full range of SNe~Ia,
including those typically labeled as peculiar and often rejected from other analyses.
We find that traditional light curve width + color standardization such as SALT2 is not sufficient.
The Twins Embedding identifies a subset of SNe~Ia including but not limited to 91T-like SNe~Ia whose
SALT2 distance estimates are biased by \saltfirstcompdiff~mag.
Standardization using the Twins Embedding also significantly decreases host-galaxy correlations. We recover a host
mass step of \hoststepfitrbtlmass~mag compared to \hoststepfitsaltmass~mag for SALT2 standardization
on the same sample of SNe~Ia. These biases in traditional standardization methods could significantly impact
future cosmology analyses if not properly taken into account.

\end{abstract}

\keywords{Type Ia supernovae --- Standard candles --- Observational cosmology}


\section{Introduction}
\label{sec:introduction}

Type~Ia supernovae (SNe~Ia) have proven to be one of the strongest probes of cosmology. SNe~Ia can
be observed out to far distances, and they can be used as standardizable candles to infer the distances to them.
The first
distance measurements with reasonably sized samples of high-redshift SNe~Ia led to the initial discovery of the
accelerating expansion of the universe \citep{riess98, perlmutter99}. Subsequent studies have now
accumulated over 1,000 spectroscopically-confirmed SNe~Ia, providing increasingly strong constraints on cosmological
parameters \citep{knop03, riess04, astier06, kowalski08, suzuki12, betoule14, scolnic18, brout19, jones19}.

\subsection{SNe~Ia as Standard Candles} \label{sec:standard_candles}

At a fixed distance, the observed peak brightnesses of SNe~Ia in the B-band have a dispersion of $\sim$0.4~mag.
To use SNe~Ia as distance estimators for cosmology, several corrections need to be applied to their
observed peak brightnesses. \citet{phillips93} showed that the peak brightnesses of SNe~Ia are tightly correlated with the
rate of decline of their light curves, commonly referred to as the ``light curve width''. \citet{riess96} and
\citet{tripp98} showed that the color of the light curve, measured as the difference between the peak brightnesses in
the B and V bands, is also highly correlated with the peak brightnesses of SNe~Ia. By combining information from the
width and color of a SN~Ia light curve, the dispersion in the corrected peak brightnesses of the
SNe~Ia is reduced to
$\sim$0.15~mag. The SALT2 model \citep{guy07, guy10, betoule14} is one of several implementing these two
corrections. SALT2 models the spectral energy distribution of SNe~Ia, and is used to estimate distances to
SNe~Ia in most modern cosmological analyses.

When distance estimates to SNe~Ia are corrected using only light curve width and color, we find that the
distance estimates are correlated with various properties of the host galaxies of SNe~Ia. These correlations are
typically modeled as a ``host step'' where SNe~Ia with a given host property below some threshold have a systematic
offset in their measured distances compared to SNe~Ia above
this threshold. SALT2-standardized distances have been shown to have ``host steps'' of
$\sim$0.1~mag when comparing SNe~Ia from host galaxies with different masses, metallicities, local colors,
local star-formation rates or global star-formation rates
\citep{kelly10, sullivan10, gupta11, dandrea11, rigault13, rigault15, rigault18, childress13, hayden13, roman18}. As galaxy properties
evolve with redshift, correlations of the peak brightness of SNe~Ia with their host-galaxy properties
would need to be well-understood to produce robust cosmological measurements.

The correlations of distance estimates to SNe~Ia with host-galaxy properties are also of interest for the
measurements of the Hubble constant with SNe~Ia. These measurements rely on the assumption that SNe~Ia in hosts
with Cepheids have similar luminosity distributions as the larger sample of SNe~Ia
\citep{riess16, riess19, rigault15}. If SNe~Ia in Cepheid-hosting galaxies have different luminosities
than the larger population SNe~Ia, then the measurements of the Hubble constant would be biased.
Host-galaxy properties are simply a proxy for some diversity of SNe~Ia
that is not captured by current standardization methods. Ideally, new standardization techniques could be
developed that identify this diversity using properties of the SNe~Ia themselves rather than properties
of their host galaxies.

Several different techniques have been proposed to improve standardization of SNe~Ia. One option is to add additional
components to a linear model like SALT2. \citet{saunders18} built a seven-component linear model (SNEMO) that is
capable of parametrizing additional diversity in the light curves of SNe~Ia compared to SALT2, and standardizes
SNe~Ia to within $0.113 \pm 0.007$~mag. Alternatively, \citet{leget20} built a
three-component linear model by first performing a PCA decomposition of the spectral features of SNe~Ia near maximum
light and then using the resulting PCA coefficients to build a linear SED model.
Another option is to incorporate additional information beyond optical light curves into
the standardization procedure. The brightnesses of SNe~Ia in the NIR are less sensitive to the intrinsic diversity of
SNe~Ia \citep{kasen06} and effects such as astrophysical dust: the corrected peak brightness of NIR light curves have
unexplained dispersions of $\sim 0.11$~mag \citep{krisciunas04, woodvasey08, mandel11, baronenugent12, burns18, stanishev18, avelino19}.

Various attempts have been made to use the spectra of SNe~Ia to standardize their brightnesses. \citet{nugent95}
showed that the ratio of the equivalent widths of the \ion{Si}{2} 5972~\AA\ and \ion{Si}{2} 6355~\AA\ features is highly
correlated with the light curve width that is typically used for standardization. \citet{bailey09} showed
that spectral flux ratios at specific wavelengths can be used to standardize SNe~Ia to within $0.125 \pm 0.011$~mag.
\citet{blondin11} and \citet{silverman12} used various spectral features directly in their
standardization and achieved dispersions of $0.143 \pm 0.020$~mag and $0.130 \pm 0.017$~mag, respectively.
\citet{nordin18} showed that various features in the U-band can be used to improve standardization.
These previous methods only used specific features of the supernova spectrum for their classification.
\citet{fakhouri15} (hereafter \citetalias{fakhouri15}) introduced a method to use all of the information
in the spectrum for standardization by identifying ``twin supernovae''.
In this methodology, the spectrum of a new supernova is compared to a large reference sample to find pairs
of spectra with very similar spectral features that are called ``twins''. Standardization using twin supernovae
resulted in a dispersion of $0.083 \pm 0.012$~mag for the sample of \citetalias{fakhouri15}, although there is some
evidence to suggest that not all twin supernovae have similar brightnesses \citep{foley20}. Furthermore,
\citetalias{fakhouri15} were only able to standardize 78\% of the SNe~Ia in their sample due to a lack of twins
for a subset of the SNe~Ia.

In a companion article (Boone et al. 2021; herefter Article~I), we showed how manifold learning can be used
to construct a parametrization of SNe~Ia using information from the pairings of twin SNe~Ia
that we call the ``Twins Embedding''.
In this work, we show how the Twins Embedding can
be used to improve distance estimates to SNe~Ia, and we discuss the biases that are present in current distance
estimation techniques.
This analysis is laid out as follows. First, we discuss the dataset that we use in Section~\ref{sec:dataset} and
summarize the methods that we developed in Article~I to build the Twins Embedding. We describe several new
standardization techniques that take advantage of this new parameter space in
Section~\ref{sec:standardization}. In Section~\ref{sec:discussion}, we compare all of these different standardization
techniques, and show the limitations of traditional SALT2 distance estimation. In the same Section, we also examine how
distances estimated with all of these different standardization techniques correlate with properties of the SNe~Ia
host galaxies. Finally, in Section~\ref{sec:conclusions}, we discuss how the results of this analysis could impact
cosmological analyses.

\section{Dataset} \label{sec:dataset}

\subsection{Overview} \label{sec:dataset_overview}

In this analysis, we use spectrophotmetric timeseries of SNe~Ia obtained by the Nearby Supernova Factory
\citep[SNfactory;][Aldering et al. 2021, in prep.]{aldering02} using the Super Nova Integral Field Spectrograph \citep[SNIFS;][]{lantz04}
on the University of Hawaii 2.2~m telescope on Mauna Kea.
This instrument collects spectra using two lenslet integral field spectrographs
\citep[IFS, ``\`a la TIGER'';][]{bacon95, bacon01}, which cover the 3200--5200~\AA\ and 5100--10000~\AA\ wavelength
ranges simultaneously. These spectrographs have a $6.^{\prime\prime}4 \times 6.^{\prime\prime}4$ field of view,
which is split into a fully-filled grid of $15 \times 15$ spatial elements. Atmospheric transmission
is monitored using a parallel imaging channel.

We reduced the spectra from this instrument using the SNfactory data reduction pipeline
\citep[][Ponder et al. 2021, in prep.]{bacon01, aldering06, scalzo10}. We calibrated the flux of each spectrum as described
in \citet{buton13}, and subtracted the host-galaxies as presented in \citet{bongard11}.
All spectra were corrected for Milky Way dust using the extinction-color relation from \citet{cardelli89} and
the dust map from \citet{schlegel98}.
Before performing this analysis, we adjusted the wavelengths and time scales of all
SNe~Ia to the restframe, and we adjusted relative brightnesses of the SNe~Ia to a common redshift.
The difference between a cosmological model and a pure Hubble law is less than 0.01~mag over
the redshift range we are considering, so our analysis is insensitive to the values of the cosmological
parameters.

In Article~I, we developed a set of techniques to process this dataset and embed
the sample of SNe~Ia into a parameter space that captures the intrinsic diversity of their spectra
at maximum light. To summarize, we first built a model of the differential phase evolution of
SNe~Ia. We used this model to estimate the spectrum of each SN~Ia at maximum light using all spectra
within five days of maximum light. We then developed a procedure that we call ``Reading Between the Lines''
(RBTL) that effectively fits for the wavelength-dependent intrinsic dispersion of SNe~Ia and uses the
regions of the spectrum with low intrinsic dispersion to estimate the peak brightness and reddening
due to dust of each SN~Ia. After this procedure, we are left with a set of dereddened spectra of SNe~Ia
at maximum light that nominally only contain intrinsic variability.

Using manifold learning techniques, we embedded these SNe~Ia into a three-dimensional
parameter space that we call the Twins Embedding. The first three components each explain a
significant fraction of the intrinsic diversity of the spectra of SNe~Ia at maximum light
(50.8, 27.0, and 11.4\% of the variance respectively), or 89.2\% with all three combined.
We find that any additional components explain less than 1\% of the remaining variance.

After applying a set of
data quality cuts based solely on measurement uncertainties and observing cadence, a total
of \nummanifoldsne\ SNe~Ia were used in the derivation of the Twins Embedding.
For each of these SNe~Ia, the RBTL algorithm provides a measurement of the SN~Ia's peak brightness and
its reddening due to dust, assuming a fiducial extinction-color relation from \citet{fitzpatrick99} with
a total-to-selective extinction ratio of $R_V = 2.8$. The reddening is parameterized by
$\Delta \tilde{A}_V$, the difference in extinction from the mean SN~Ia template. The validity of this fiducial extinction-color
relation will be discussed in Section~\ref{sec:rbtl_standardization} (although the derivation
of the Twins Embedding is insensitive to the details of this relation).

The spectrophotometric spectra of SNe~Ia were shifted to a common redshift before applying the
RBTL algorithm. Hence by comparing the brightness estimate from the RBTL algorithm to the mean peak brightness
across the entire sample of SNe~Ia, we obtain ``RBTL magnitude residuals'' that describe how bright
a SN~Ia is relative to the rest of the sample. For a perfect standardization technique, these
magnitude residuals would all be zero. Note that the RBTL magnitude residuals are
estimated using only a single spectrum at maximum light rather than a full light curve, and can be
interpreted as our ability to standardize SNe~Ia with a single spectrophotometric spectrum near maximum light.
The RBTL algorithm only produces an estimate of the peak brightness of a SN~Ia, but does not take into
account any correlations between the peak brightness and other properties of the SN~Ia such as spectral
features or light curve width. In the rest of this article, we will discuss how the Twins Embedding can be
used to standardize the magnitudes residuals derived from the RBTL algorithm.

\subsection{Sample of Type~Ia Supernovae for Standardization Analyses} \label{sec:magnitude_requirements}

There are several sources of uncertainty in the magnitude residuals that are not due to intrinsic
differences between SNe~Ia. When shifting all of our observations of SNe~Ia to a common
redshift, we relied on the observed redshifts of each SN~Ia. Any uncertainties on these observed redshifts
will propagate into our magnitude residuals. \NumToName{\numsnredshift} of the \nummanifoldsne\ SNe~Ia
in the Twins Embedding have redshifts derived from the supernova spectrum since their
host galaxies are too faint to yield a spectrum. The uncertainties on these redshifts are $\sim$0.004,
which contributes at least 0.10~mag to the magnitude residual uncertainties. We therefore reject these SNe~Ia
from our standardization analyses. These \numtoname{\numsnredshift} SNe~Ia are a small fraction of our
low-mass hosts \citep{childress13}, so we can reject them without biasing our analyses.

Similarly, peculiar velocities of the supernova host galaxies will introduce an additional contribution
to the measured magnitude residuals. Assuming a typical dispersion for the peculiar velocities of 300~km/s
\citep{davis11}, the introduced dispersion in brightness becomes comparable to the dispersion in brightness
of SNe~Ia around $z=0.02$ where 0.11~mag of dispersion is introduced. We reject
\numlowredshift\ SNe~Ia that are at redshifts lower than 0.02 out of the \nummanifoldsne\ SNe~Ia in the
Twins Embedding. For redshifts above 0.02, the contribution from peculiar velocities is still significant,
so we propagate the implied uncertainties on the magnitude residuals in our analyses. For the selected 
sample of SNe~Ia used in this analysis, peculiar velocities contribute \pecvelcontribution~mag to the dispersion of
the magnitude residuals.

Finally, dust in the host galaxy of a SN~Ia can significantly dim the observed brightness of the SN~Ia.
While the mean total-to-selective extinction ratio for dust was measured to be $R_V = 2.8 \pm 0.3$ in
\citet{chotard11}, there are examples of SNe~Ia with very different values of $R_V$. For example, the
extinction of the highly reddened SN2014J has been suggested to be
either interstellar dust with $R_V = 1.4 \pm 0.1$ \citep{amanullah14} or
circumstellar dust with a similar effect on the observed spectrum \citep{foley14}.
Studies of large samples of SNe~Ia have confirmed that there is increased dispersion in the
magnitude residuals of highly reddened SNe~Ia \citep{brout20}.
For our analysis, if an incorrect value of $R_V$ is assumed for a SN~Ia, then the primary effect is that a
constant offset is added to the SN~Ia's magnitude residual. For SN2014J, the
difference in assuming a dust law with $R_V = 2.8$ instead of $R_V = 1.4$ would introduce an offset of
$\sim$1 mag into the observed magnitude residual. To avoid having differences in dust properties affect our analysis
of magnitude residuals, we choose to reject any supernovae with a measured RBTL $\Delta \tilde{A}_{V} > 0.5$
from our standardization analyses.
For a supernova with $\Delta \tilde{A}_{V} = 0.5$, a 20\% difference in $R_V$ would introduce a
$\sim$0.1 mag offset into its magnitude residual.
The distribution of $\Delta \tilde{A}_{V}$ is constructed to have a median of zero, and this
selection cut is approximately equivalent to a selection cut of SALT2 $c > 0.2$.
\NumToName{\numhighav} supernovae have measured $\Delta \tilde{A}_{V}$ values larger
than this threshold out of the \nummanifoldsne\ in the Twins Embedding.

For the SNe~Ia passing these selection criteria, variation in $R_V$ will have a less than 0.01~mag
effect on the observed spectra beyond a flat offset in the magnitude residual. This variation is too small
to be captured by direct measurement of the spectrum, and will not affect the location of a SN~Ia in
the Twins Embedding. In Section \ref{sec:rbtl_standardization}, we will show how the mean $R_V$ value can be
recovered by looking at correlations between $\Delta \tilde{A}_V$ and magnitude residuals.

We find that a total of \nummagsne\ SNe~Ia pass all of the selection requirements
for their magnitude residuals to be expected to have low contributions from non-intrinsic sources.

\subsection{SALT2 Fits} \label{sec:salt_fits}

For comparison purposes, we fit the SALT2 light curve model \citep{betoule14} to
all of the light curves in our sample. For the fit,
we use synthetic photometry in the SNfactory \Bsnf, \Vsnf, and \Rsnf bands defined as tophat filters
with wavelength ranges shown in Table~\ref{tab:snfactory_bands}.
We reject unreliable SALT2 fits following a procedure similar to \citet{guy10}. First, we impose phase
coverage criteria by requiring at least five measurements total at different restframe times
$t$ relative to maximum light, at least four measurements 
satisfying $-10 < t < 35$ days, at least one measurement with $-10 < t < 7$
days, at least one measurement
with $7 < t < 20$ days and measurements in at least two different synthetic filters with $-8 < t < 10$ days.
We then require that the normalized median absolute deviation (NMAD) of the SALT2 model residuals
be less than 0.12~mag and that no more than 20\% of the SALT2 model residuals have an amplitude
of more than 0.2~mag. Note that these selection criteria reject some SNe~Ia that are not
well-modeled by SALT2, so the subsample with SALT2 fits does not cover the full parameter space of SNe~Ia.
In particular, we find that seven SNe~Ia that are 91bg-like or similar to 91bg-like SNe~Ia
are rejected, and the two 02cx-like SNe~Ia in our sample are rejected. See \citet{lin20submitted}
for details on these subclasses of SNe~Ia.  A total of \numsaltsne\ of the \nummanifoldsne\ SNe~Ia in the Twins
Embedding have valid SALT2 fits.

\begin{deluxetable}{ll}
\tablecaption{
    SNfactory filter definitions used for synthetic photometry. These are tophat filters with perfect
    transmission in the given wavelength range and no throughput outside of that range.
}
\label{tab:snfactory_bands}
\tablehead{
    \colhead{Band Name} & \colhead{Wavelength Range (\AA)}
}
\startdata
    \Usnf & 3300--4102 \\
    \Bsnf & 4102--5100 \\
    \Vsnf & 5200--6289 \\
    \Rsnf & 6289--7607 \\
    \Isnf & 7607--9200 \\
\enddata
\end{deluxetable}

\subsection{Blinding of the Analysis} \label{sec:blinding}

To avoid tuning our analysis to optimize the distribution of the magnitude residuals,
we performed this analysis ``blinded'': we split the dataset into a training and a validation set,
and we optimized our analysis while only examining the training set, leaving the validation set to only be
examined when the analysis and selection requirements were complete and well-understood for the training set.

We split the SNe~Ia with valid SALT2 fits evenly into ``training'' and
``validation'' subsets while attempting to match the distribution of redshift and SALT2 parameters across the
two subsets. The subset of SNe~Ia that does not have valid SALT2 fits is small (\numbadsalt\ of
the \nummanifoldsne\ in the Twins Embedding), and contains both
unusual SNe~Ia and those whose SALT2 fits are impacted by spectra with relatively large
instrumental uncertainties.
Spectra with large instrumental uncertainties could
highly influence our analysis, so we included all of the SNe~Ia without valid SALT2 fits in the training
set. We used this combined training subset to develop our model and to decide upon a set of selection
requirements that removes spectra with large instrumental uncertainties. The SNe~Ia in the ``validation'' subset
were only examined after we had settled on the final model and selection requirements, and no further changes
were made to the analysis after ``unblinding'' the validation subset.
The combined training subset contains \numsnftraincombined\ of the \nummanifoldsne\ SNe~Ia in the Twins
Embedding and the validation subset contains \numsnfvalid\ SNe~Ia. The blinding was implemented by immediately
deleting all of the magnitudes estimated for SNe~Ia in the ``validation'' subset as soon as the RBTL model had been run,
so that it was impossible to accidentally unblind the distributions of the magnitude residuals. 
Prior to unblinding, we decided that for the baseline result of each analysis we would report the results
when running the analysis only on the training set, only on the validation set, and on the full dataset.

As we are reporting statistics on distributions of magnitude residuals, one potential concern is that a single
unusual SN~Ia with a very large magnitude residual could highly skew the measured statistics. To address this
concern, we decided before unblinding to calculate both the RMS and the NMAD of the magnitude residuals for all of our analyses.
We attempted to come up with a robust set of selection criteria in Section~\ref{sec:magnitude_requirements} before unblinding.
However, it is possible that our selection criteria could accidentally include a small number of SNe~Ia in the validation
set with large instrumental uncertainties that inflate their measured magnitude residuals. We agreed in advance that if
this were to happen,
we would decide post-unblinding to report statistics both on the full sample and on the sample with that subset removed.
We did not see any evidence of such a subset after unblinding, and all of our
results on the validation set are consistent with the results on the training set. Unless otherwise noted,
reported numbers in the text and figures are for the full dataset.

The results of all of the selection requirements described in this Section are summarized in Table~\ref{tab:selection_requirements}.

\begin{deluxetable*}{ll}
\tablecaption{Summary of the sample selection requirements.}
\label{tab:selection_requirements}
\tablehead{
    \colhead{Selection Requirement} & \colhead{Number of SNe~Ia} \\
    & \colhead{Passing Requirement}
}
\startdata
    \textbf{General selection requirements} & \\
SNe Ia Included in Twins Embedding (Paper I)                & 173 \\
\hline
\textbf{Standardization of near-maximum spectra} & \\
\textbf{(Section~\ref{sec:magnitude_requirements})} & \\
Host galaxy redshift available                      & 168 \\
Host galaxy redshift above 0.02                     & 144 \\
RBTL $\Delta \tilde{A}_V$ < 0.5 mag                            & 134 \\
Blinded training subsample                          & 72 \\
Validation subsample                                & 62 \\
\hline
\textbf{Comparisons to SALT2 standardization} & \\
\textbf{(Section~\ref{sec:salt2_standardization})} & \\
Passes SALT2 selection requirements                 & 155 \\
Passes host galaxy redshift and color requirements  & 127 \\
Blinded training subsample                          & 66 \\
Validation subsample                                & 61 \\

\enddata
\end{deluxetable*}

\section{Standardizing the Magnitude Residuals of SNe~Ia} \label{sec:standardization}

\subsection{SALT2 Standardization} \label{sec:salt2_standardization}

To set a performance baseline, we first standardize the magnitude residuals of our
sample of SNe~Ia with conventional SALT2 standardization. We use SALT2 fits to the light curve
to determine the peak brightness of each SN~Ia, and we apply linear corrections for the light
curve width and color to estimate the distance to each SN~Ia. From the SALT2 fits described in
Section~\ref{sec:salt_fits}, we obtain the SALT2
parameters $x_{1, i}$ and $c_i$ of each SN~Ia along with the observed peak B-band magnitude $m_{\textrm{B},i}$.
Since we first shifted all of our observations to a common redshift, $m_{\textrm{B},i}$ is
effectively the brightness of a SN~Ia at an arbitrary fixed distance common to all of the
SNe~Ia in the sample. Given a set of standardization
parameters $\alpha$ and $\beta$ along with an arbitrary reference magnitude $m_{\textrm{ref}}$, we
calculate the SALT2 magnitude residuals $m_{\textrm{res},i}$ for each supernova as:
\begin{align}
    m_{\textrm{res},i} = m_{\textrm{B},i} - m_{\textrm{ref}} + \alpha \times x_{1,i} - \beta \times c_i
\end{align}

We estimate the uncertainty on $m_{\textrm{res},i}$ for each supernova by propagating the uncertainties from the
SALT2 fits for $x_{1,i}$, $c_i$ and $m_{\textrm{B},i}$. Additionally, we include the contribution of peculiar velocities
$\sigma_{\textrm{p.v.},i}$ assuming a dispersion of 300~km/s as described in Section~\ref{sec:magnitude_requirements}.
The SALT2 model is known to not explain all of the dispersion of SNe~Ia, so the uncertainty on this magnitude
residual must include an unexplained dispersion term $\sigma_{u}$. The final uncertainty model for
$m_{\textrm{res},i}$ is:
\begin{align}
    \sigma_{m_{\textrm{res}},i}^2 = & \sigma_{m_\textrm{B},i}^2 + \sigma_{\textrm{p.v.},i}^2 + \sigma_{u}^2 + \alpha^2 \sigma_{x_1,i}^2 + \beta^2 \sigma_{c,i}^2 \\
    & + 2\alpha\ \mathrm{Cov}[m_{\textrm{B},i} , x_1] - 2\beta\ \mathrm{Cov}[m_{\textrm{B},i}, c] \nonumber \\
    & - 2\alpha\beta\ \mathrm{Cov}[x_1,c] \nonumber
\end{align}

Given a set of parameters for the above equations, we define the following weighted RMS (WRMS) and $\chi^2$ per
degree-of-freedom ($\chi^2 / DoF$):
\begin{align}
    \textrm{WRMS} = \sqrt{\frac{\sum_i^N m_{\textrm{res},i}^2 / \sigma_{m_{\textrm{res}},i}^2}{\sum_i^N 1 / \sigma_{m_{\textrm{res}},i}^2}}
    \label{eq:salt_wrms}
\end{align}
\begin{align}
    \chi^2 / DoF = \frac1{N-4} \sum_i^N \frac{m_{\textrm{res},i}^2}{\sigma_{m_{\textrm{res}},i}^2}
    \label{eq:salt_chi2}
\end{align}
where $N$ is the total number of supernovae in the sample. We iteratively minimize these two equations to determine
the values of $\alpha$, $\beta$, $m_{\textrm{ref}}$ and $\sigma_{u}$. First, we set $\sigma_{u}$ to a guess of 0.1~mag. We then
minimize the WRMS in Equation~\ref{eq:salt_wrms} to determine the optimal parameters for $\alpha$, $\beta$, and $m_{\textrm{ref}}$.
Given the fitted values of these parameters, we determine the value of $\sigma_{u}$ that sets the $\chi^2/DoF$ in
Equation~\ref{eq:salt_chi2} to 1. We repeat these two fits until the parameter values converge.

The results of this SALT2 standardization procedure are summarized in Table~\ref{tab:salt_parameters}. As SALT2 fits
to the SNfactory dataset have been studied in many previous analyses, we did not blind these numbers. The uncertainties
on these measurements, and all other uncertainties on measurements of dispersion in this analysis, are calculated
using bootstrapping \citep{efron79}. We find a WRMS of \saltparamwrms~mag for this sample using SALT2.
Because the unexplained dispersion $\sigma_u$ is large compared to the typical measurement and peculiar
velocity uncertainties, there is little difference in the total uncertainties
$\sigma_{m_{\textrm{res}},i}$ between SNe~Ia, with values ranging between
\saltparammindisp~mag to \saltparammaxdisp~mag. Hence the unweighted RMS of \saltparamrms~mag is nearly identical
to the WRMS. To compare with other standardization techniques and avoid having different weights for different techniques,
we choose to use the unweighted RMS in further analyses. Interestingly, we also find that this distribution
has a tight core with wider wings compared to a Gaussian distribution, as seen in the low NMAD of only \saltparamnmad~mag
for this sample.

\begin{deluxetable}{ll}
\tablecaption{
    SALT2 standardization parameters fit using the procedure described in Section~\ref{sec:salt2_standardization}.
}
\label{tab:salt_parameters}
\tablehead{
    \colhead{Parameter} & \colhead{Value}
}
\startdata
    $x_1$ correction ($\alpha$) & \saltparamalpha \\
    Color correction ($\beta$) & \saltparambeta \\
    Unexplained dispersion ($\sigma_{u}$) & \saltparamsigmaint~mag \\
    Weighted RMS of $m_{\textrm{res},i}$ & \saltparamwrms~mag \\
    Unweighted RMS of $m_{\textrm{res},i}$ & \saltparamrms~mag \\
    NMAD of $m_{\textrm{res},i}$ & \saltparamnmad~mag \\
\enddata
\end{deluxetable}

\subsection{Raw RBTL Magnitude Residuals} \label{sec:raw_rbtl}

With the selection criteria from Section~\ref{sec:magnitude_requirements} applied, we find that the RBTL
magnitude residuals have a dispersion with an unweighted RMS of \rawrbtlmagstd~mag
and an NMAD of \rawrbtlmagnmad~mag
for the validation set. Note that these magnitude residuals have been corrected
using a baseline extinction-color relation (correcting for both dust and any intrinsic color that has
a similar functional form), but they have not been corrected for any other intrinsic properties
that do not affect the intrinsic color of SNe~Ia.

\subsection{Standardizing the RBTL Magnitude Residuals with the Twins Embedding} \label{sec:rbtl_standardization}

The RBTL algorithm provides a robust estimate of the brightness and extinction of the spectrum
of a SN~Ia. However, if
there is intrinsic diversity that affects the spectrum in a similar manner to a brightness difference
or an extinction, then it
will be confused as extrinsic diversity by the RBTL algorithm. This means that at some level the RBTL
magnitude residuals will include contributions from intrinsic diversity. Assuming that the intrinsic diversity that affects
the brightness of the supernova also affects other intrinsic properties of the spectra, such as the
features seen in the spectra of SNe~Ia, then we can use the intrinsic diversity measured from the spectra
to remove the intrinsic contributions to the magnitude residuals. This procedure is similar in concept
to what is done
with light curve fitters such as SALT2: the initial distance estimate comes from the observed brightness of a SN~Ia
in the B-band. However, this distance estimate contains contributions from the intrinsic diversity (measured with $x_1$),
and a correction is applied to the original distance estimate to remove these contributions.

\begin{figure*}
    \epsscale{0.8}
    \plotone{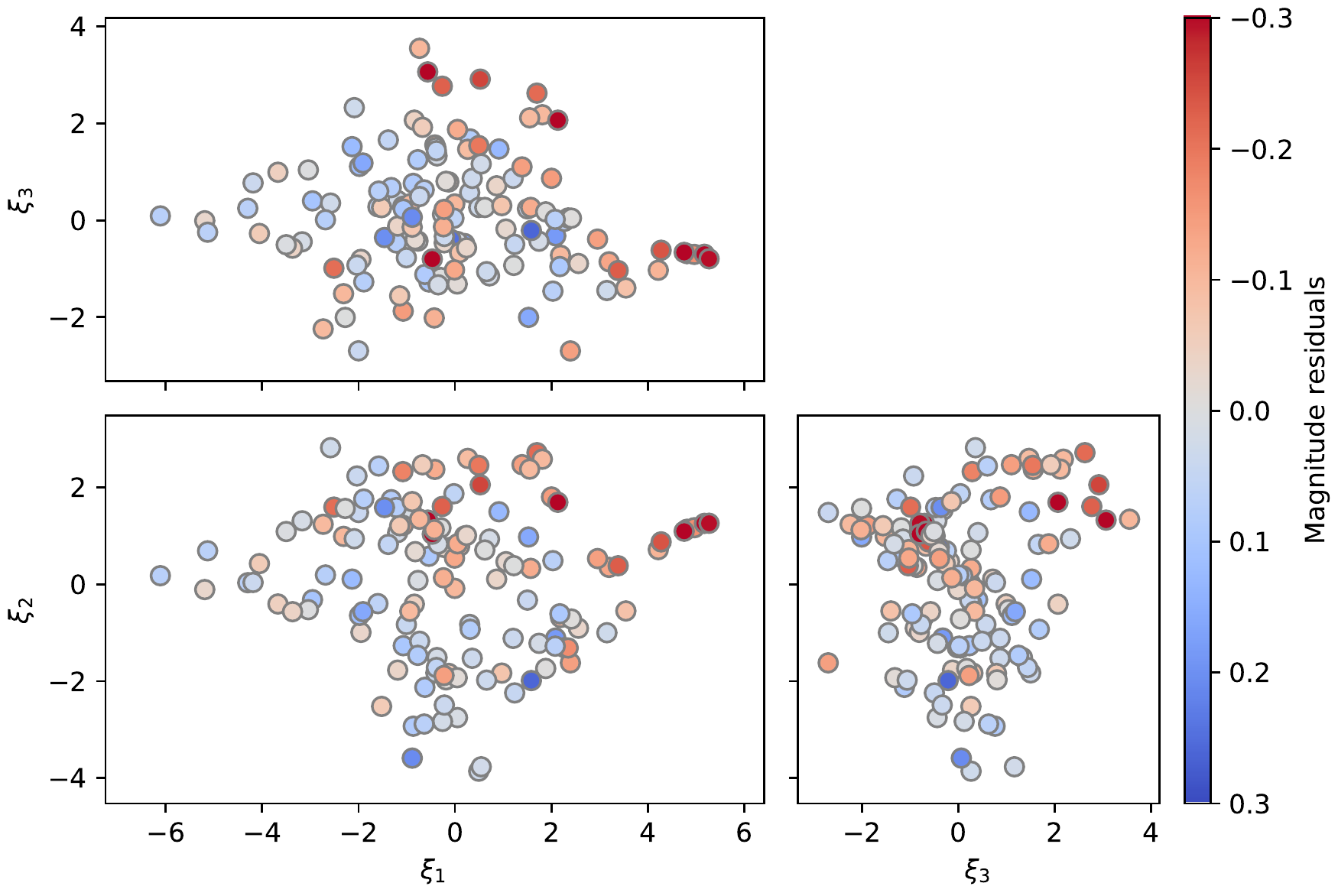}
    \caption{
        RBTL magnitude residuals as a function of the Twins Embedding coordinates $\xi_i$. Each point represents
        a single SN~Ia. The data points are colored according to their RBTL magnitude residuals, as labeled in the
        colorbar. Each panel shows a different 2D-projection of the Twins Embedding.
    }
    \label{fig:rbtl_mags_components}
\end{figure*}

The Twins Embedding is a parametrization of the intrinsic diversity of SNe~Ia. In Figure~\ref{fig:rbtl_mags_components},
we show the RBTL magnitude residuals as a function of the location of SNe~Ia in the Twins Embedding, with
two-dimensional projections for each of the three dimensions of the Twins Embedding.
There are visible trends
in the magnitude residuals as a function of each of the components: SNe~Ia that are nearby in the
Twins Embedding tend to have similar RBTL magnitude residuals, indicating that the RBTL brightness
estimates indeed contain contributions from the intrinsic diversity.

The Twins Embedding was designed to capture highly nonlinear features, so there is no reason to expect that
the resulting components will be linearly correlated with the peak brightnesses of SNe~Ia. Linear standardization
as is traditionally done for light curve fitters such as SALT2 is unlikely to be sufficient. Instead, we choose
to use Gaussian Process (GP) regression to model the magnitude residuals of SNe~Ia over the
Twins Embedding. GP regression effectively generates a prediction of the magnitude residuals at a target location in
the Twins Embedding using the observed magnitude residuals in a region around that target location. It is
also able to propagate uncertainties on the predictions.
For an introduction to GP regression, see Appendix~\ref{sec:gaussianprocess}.

When calculating the RBTL magnitude residuals, as described in Section \ref{sec:magnitude_requirements}, we chose
to use a fiducial extinction-color relation with $R_V = 2.8$. If the wrong mean value of $R_V$ is chosen,
then the only observable effect for this analysis is that a correlation is introduced between the
measured extinction and the magnitude residuals. We account for this potential difference in $R_V$ by using a mean
function in our GP that contains a correction term $\omega$ that is linear in $\Delta \tilde{A}_V$ along with an arbitrary reference
magnitude $m_{\textrm{ref}}$.
For each SN~Ia, the magnitude residuals contain uncorrelated measurement uncertainties due to the peculiar velocity of the SN~Ia's
host galaxy $\sigma_{\textrm{p.v.},i}$ (described in Section~\ref{sec:magnitude_requirements}) and an unexplained
residual dispersion $\sigma_{u}$ whose value will be determined in the fit. Measurement uncertainties from the RBTL
algorithm are negligible for the high signal-to-noise observations used in this analysis. Any systematic
uncertainties from our modeling procedures will be captured by $\sigma_{u}$.

Finally, we use a Mat\'ern 3/2 kernel (see Appendix~\ref{sec:gaussianprocess}) to describe the correlation of the magnitude residuals of SNe~Ia across the
Twins Embedding. This kernel has two parameters: a length scale $l$ that determines the distances in the Twins Embedding
over which the magnitude residuals of different SNe~Ia are coherent, and an amplitude $A$ that sets the size of
those coherent variations. The Twins Embedding was designed so that the distance between two SNe~Ia in
the Twins Embedding is proportional to the size of the differences between their spectra.
Hence, we choose to use a three-dimensional Mat\'ern kernel with a single length scale. The full GP model is
then:
\begin{align}
    \vec{m}_{\textrm{RBTL}} \sim \mathcal{GP} \Bigl( & m_{\textrm{ref}} + \omega \Delta \vec{\tilde{A}}_V, \nonumber \\
                                                    & \mathbf{I} \cdot (\vec{\sigma}_\textrm{p.v.}^2 + \sigma_u^2) + K_{3/2} (\vec{\xi}, \vec{\xi}; A, l) \Bigl)
\end{align}
where $\vec{m}_{\textrm{RBTL}}$ are the RBTL magnitude residuals and $\mathbf{I}$ is the identity matrix.

We implement this GP model using the \texttt{George} package \citep{ambikasaran15}. The full model
has a total of five parameters: $m_{\textrm{ref}}$, $\omega$, $\sigma_u$, $A$, and $l$. We fit this model to the
sample of SNe~Ia described in Section~\ref{sec:magnitude_requirements} optimizing the maximum likelihood.
The results of this fit are shown in Table~\ref{tab:gp_standardization_parameters}.

\begin{deluxetable}{ll}
\tablecaption{
    Parameters for the GP standardization model described in Section~\ref{sec:rbtl_standardization}.
}
\label{tab:gp_standardization_parameters}
\tablehead{
    \colhead{Parameter} & \colhead{Value}
}
\startdata
    Extinction $R_V$ & \rbtlgprv \\
    GP kernel amplitude ($A$) & \rbtlgpkernelamp~mag \\
    GP kernel length scale ($l$) & \rbtlgpkernellengthscale \\
    Unexplained dispersion ($\sigma_u$) & \rbtlgpintdisp~mag \\
    RMS of $m_{\textrm{corr.},i}$ & \rbtlgprms~mag \\
    NMAD of $m_{\textrm{corr.},i}$ & \rbtlgpnmad~mag \\
    \hline
\enddata
\end{deluxetable}

Given the fitted extinction correction $\omega$, our estimate of the true extinction-color relation is then the
fiducial one with that correction added as a zeropoint offset. We estimate the true value of $R_V$ by solving
for the extinction-color relation from \citet{fitzpatrick99} that best matches this estimated true
extinction-color relation. We find that our measurements prefer
an $R_V$ value of \rbtlgprv. We verified that different
choices of fiducial extinction-color relations result in similar recovered $R_V$ values when rerunning the entire analysis.
Note that we have rejected SNe~Ia with $\Delta \tilde{A}_{V} > 0.5$ from this analysis, so our measurement of $R_V$
is made using only SNe~Ia with relatively low extinction. Previous measurements of $R_V$ from optical spectra include
$R_V = 2.8 \pm 0.3$ from \citet{chotard11} and $R_V = 2.6 \pm 0.5$ from \citet{leget20}. \citet{mandel20} found 
$R_V = 2.9 \pm 0.2$ from measurements of SNe~Ia with infrared photometry. Our results are compatible with, but slightly
lower than, these previous measurements of $R_V$. The fact that our sample is restricted to low-extinction SNe~Ia
makes direct comparisons challenging.

\begin{figure*}
    \epsscale{0.8}
    \plotone{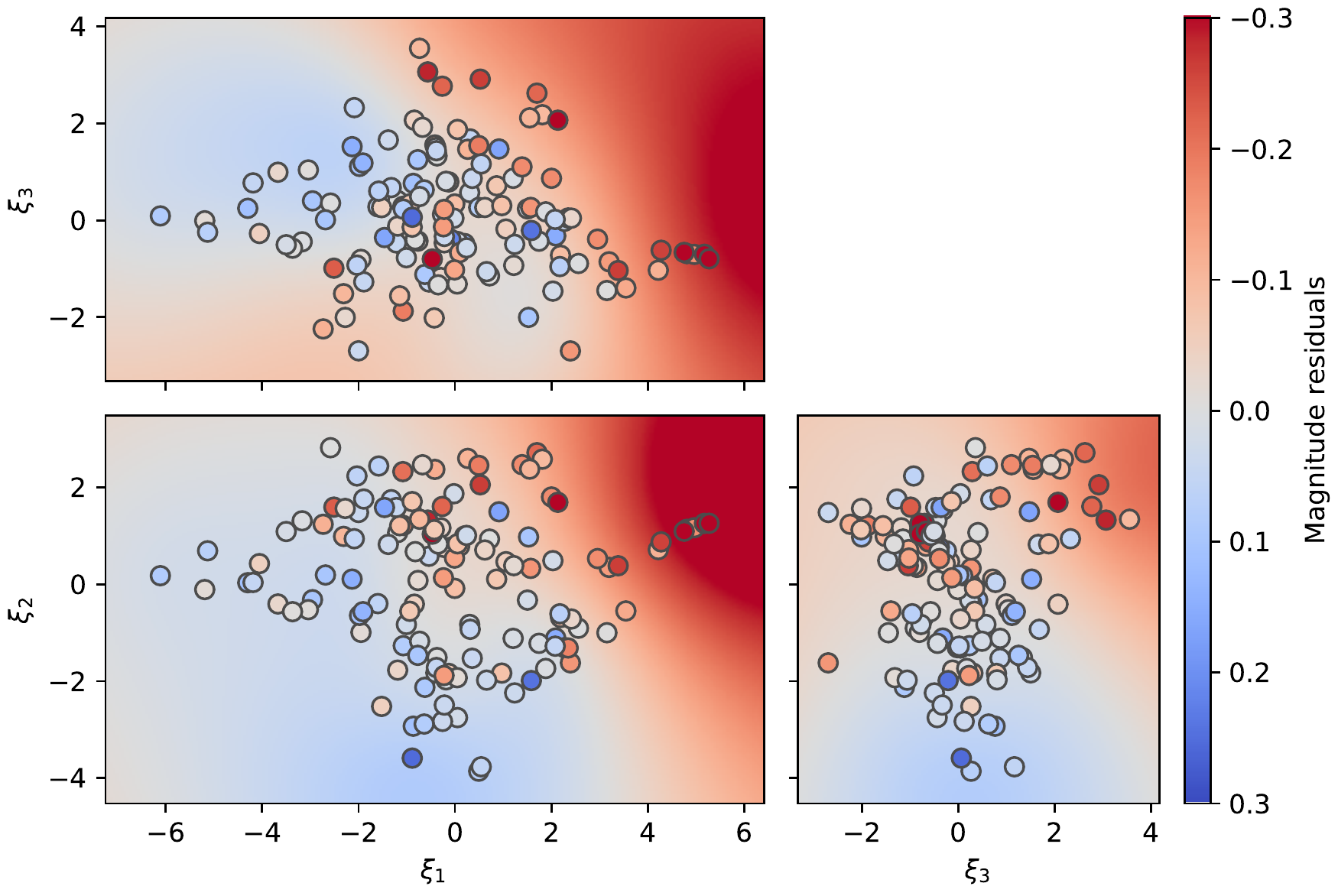}
    \caption{
        RBTL magnitude residuals and GP predictions as a function of the Twins Embedding coordinates $\xi_i$.
        Each point represents a single SN~Ia, colored according to its RBTL magnitude residual. Each
        panel shows a 2D-projection of the Twins Embedding. The smooth color in the background indicates the amplitude of the
        GP prediction for each value of the two displayed components while the remaining component is fixed to zero.
    }
    \label{fig:rbtl_gp_components}
\end{figure*}

The recovered unexplained dispersion of \rbtlgpintdisp~mag is significantly lower than what is typically found for
light curve fitters that only rely on light curve width and color, and is consistent with the results of \citetalias{fakhouri15}
for pairs of twin supernovae, as will be discussed in Section~\ref{sec:standardization_comparison}.
To understand the behavior of the GP, we show the GP predictions after applying the updated $R_V$ extinction correction in
Figure~\ref{fig:rbtl_gp_components}. In these figures, we hold one
component value fixed to zero, and show the effects of varying the other two components over the Twins Embedding.
Note that these figures only show slices through the GP predictions, and the predictions for individual SNe~Ia
will also include information from the remaining component that cannot be plotted in only two dimensions.
Qualitatively, the GP appears to be able to capture the nonlinear variation in the RBTL magnitude residuals across
the Twins Embedding.

\begin{figure*}
    \epsscale{0.8}
    \plotone{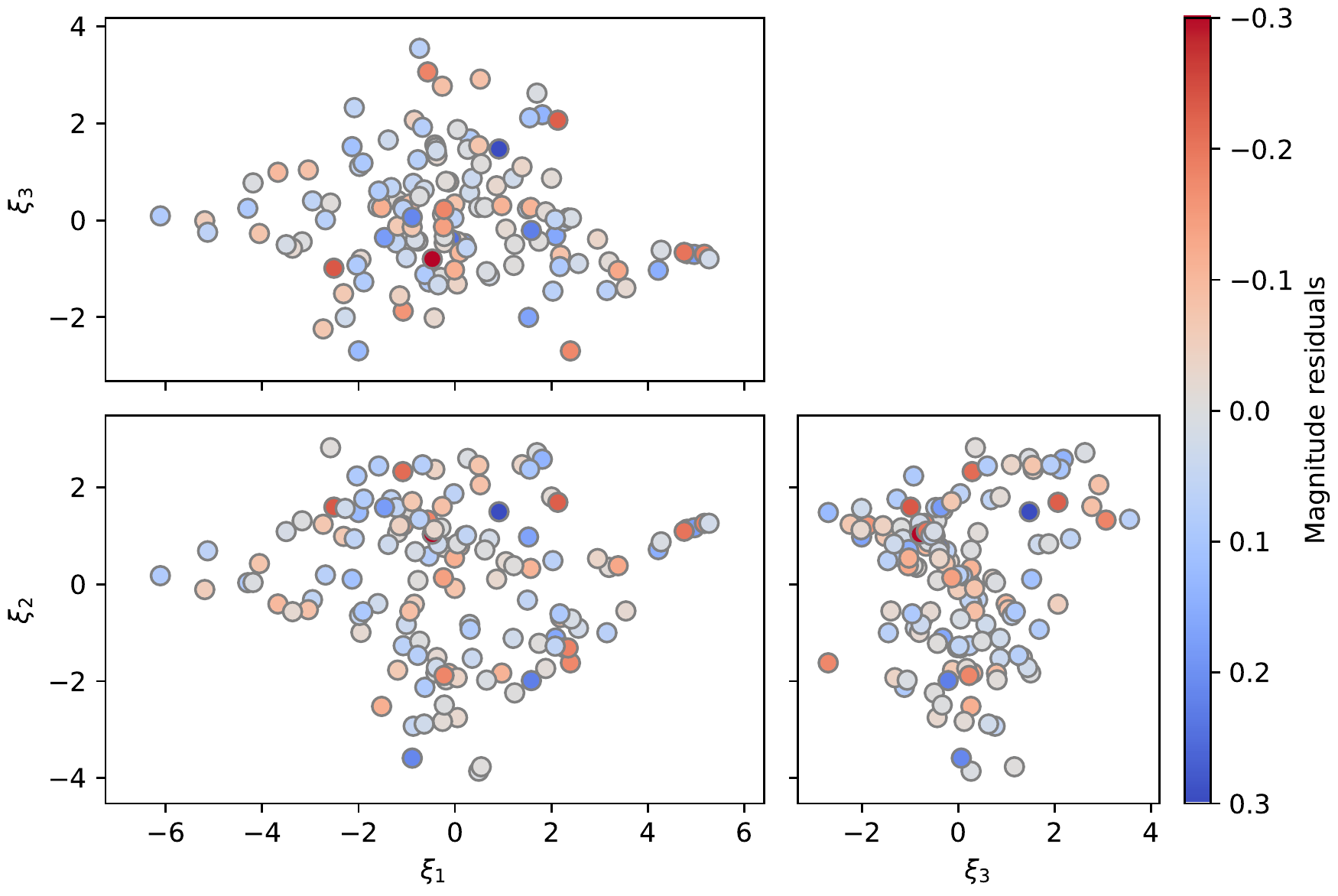}
    \caption{
        GP-corrected RBTL magnitude residuals $m_{\textrm{corr.},i}$ as a function of the Twins
        Embedding coordinates $\xi_i$.
        Each point represents a single SN~Ia. The data points are colored according to their GP-corrected
        RBTL magnitude residuals $m_{\textrm{corr.},i}$, as labeled in the
        colorbar. Each panel shows a different 2D-projection of the Twins Embedding.
        There is little evidence of correlation between the magnitude residuals
        of SNe~Ia that are nearby in the Twins Embedding, and the remaining observed dispersion
        is largely driven by peculiar velocity uncertainties.
    }
    \label{fig:rbtl_gp_residuals}
\end{figure*}

We can use the GP to predict what the RBTL magnitude residuals should be for any location in the Twins Embedding,
and use these predictions to correct the raw RBTL magnitude residuals.
We estimate the magnitude residual of each SN~Ia with the GP using individual ``leave-one-out'' predictions for each
SN~Ia where the GP is conditioned on all SNe~Ia except the supernova of interest. This ensures that the measured
brightness of that supernova itself cannot contribute to its own predictions. Using these predictions, we calculate
GP-corrected magnitude residuals $m_{\textrm{corr.},i}$ for each SN~Ia in our sample as the difference between the
raw RBTL magnitude residual and the GP prediction. In Figure~\ref{fig:rbtl_gp_residuals}
we show $m_{\textrm{corr.},i}$ as a function of the location of SNe~Ia in the Twins Embedding.
In contrast to Figure~\ref{fig:rbtl_mags_components}, where we saw strong trends in the raw RBTL magnitude residuals
across the twins embedding, there are no visible correlations in $m_{\textrm{corr.},i}$ for SNe~Ia that are in
similar locations in the Twins Embedding.
For the full dataset, we find that the GP-corrected magnitude
residuals $m_{\textrm{corr.},i}$ have a dispersion in brightness with an unweighted RMS of \rbtlgprms~mag and
an NMAD of \rbtlgpnmad~mag. These dispersions are larger than the unexplained dispersion of \rbtlgpintdisp~mag
because they contain contributions from both peculiar velocities and uncertainties in the GP predictions
for SNe~Ia in sparsely populated regions of the Twins Embedding.

\subsection{Standardizing the SALT2 Magnitude Residuals with the Twins Embedding} \label{sec:salt2_isomap_standardization}

We investigate a hybrid standardization method where we use the peak brightness and color from a SALT2 light
curve fit, and the Twins Embedding to parametrize the intrinsic variation of SNe~Ia instead of the SALT2
$x_1$ parameter. While the RBTL method requires that the input spectra be spectrophotometric to obtain accurate
estimates of the brightness and extinction of each SN~Ia, it may be possible to locate a SN~Ia in
the Twins Embedding using slit-based spectrographs or other forms of spectroscopy that are not flux-calibrated since most of
the information is in local spectral feature variation. External estimates of the brightness and color, from
SALT2 fits to photometry for example, could then be corrected using the Twins Embedding.

To test whether this method of standardization might be effective, we ran the GP standardization procedure described
in Section~\ref{sec:rbtl_standardization} using the Twins Embedding, but using the raw SALT2 magnitude residuals
and colors (uncorrected for $x_1$) instead of the RBTL magnitude residuals and extinctions. As for the RBTL analysis, we include
a linear term to correct for the SALT2 color, which in this context is equivalent to the $\beta$ parameter in traditional SALT2
analyses. The results of the GP fit for this SALT2 + Twins Embedding model are summarized in
Table~\ref{tab:salt2_isomap_standardization_parameters}.

\begin{deluxetable}{ll}
\tablecaption{
    Parameters for the hybrid SALT2 + Twins Embedding standardization model described in
    Section~\ref{sec:salt2_isomap_standardization}.
}
\label{tab:salt2_isomap_standardization_parameters}
\tablehead{\colhead{Parameter} & \colhead{Value}}
\startdata
    Color correction ($\beta$) & \saltgpcolor \\
    GP kernel amplitude ($A$) & \saltgpkernelamp~mag \\
    GP kernel length scale ($l$) & \saltgpkernellengthscale \\
    Unexplained dispersion ($\sigma_u$) & \saltgpintdisp~mag \\
    RMS of $m_{\textrm{corr.},i}$ & \saltgprms~mag \\
    NMAD of $m_{\textrm{corr.},i}$ & \saltgpnmad~mag \\
\enddata
\end{deluxetable}

\begin{figure*}
    \epsscale{0.8}
    \plotone{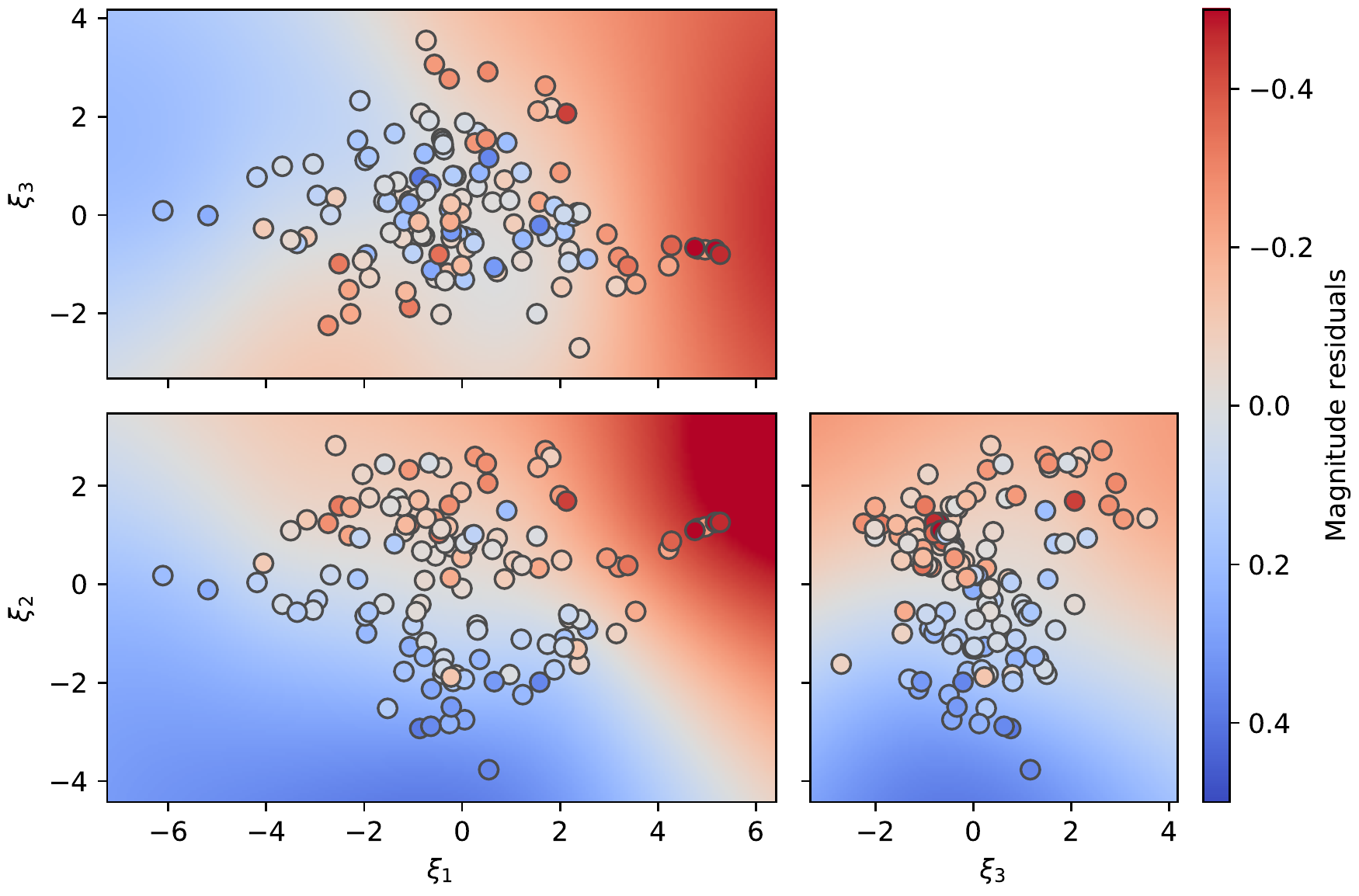}
    \caption{
        SALT2 magnitude residuals and GP predictions as a function of the Twins Embedding coordinates $\xi_i$.
        Each point represents a single SN~Ia, colored according to its SALT2 magnitude residual after applying the
        color correction. In each panel,
        the smooth color in the background indicates the amplitude of the GP prediction for each value of the two displayed
        components while the remaining component is fixed to zero.
    }
    \label{fig:salt_gp_components}
\end{figure*}

As in Section~\ref{sec:rbtl_standardization}, we estimate the SALT2 + Twins Embedding GP brightness
for each SN~Ia using leave-one-out predictions. When raw SALT2
magnitude residuals are standardized using the Twins Embedding, the standardization performance is significantly
better than standardizing only on SALT2 $x_1$. The resulting unweighted RMS of the SALT2 + Twins Embedding standardized
magnitude residuals is \saltgprms~mag compared to \saltparamrms~mag for SALT2 with traditional $x_1$ standardization
for the same set of SNe~Ia. Interestingly, the NMAD values of the standardized magnitude residuals are
nearly identical
for these two analyses (\saltgpnmad~mag compared to \saltparamnmad~mag), implying that most of this improvement
comes from improving standardization of SNe~Ia in the tails of the distribution rather than the core.
To illustrate the effect of the Twins Embedding for standardization of raw SALT2 magnitude residuals, we
show the GP predictions for the SALT2 + Twins Embedding standardization analysis
in Figure~\ref{fig:salt_gp_components} along with the color-corrected magnitude residuals for all of the SNe~Ia
in our sample.

\begin{figure*}
    \epsscale{0.8}
    \plotone{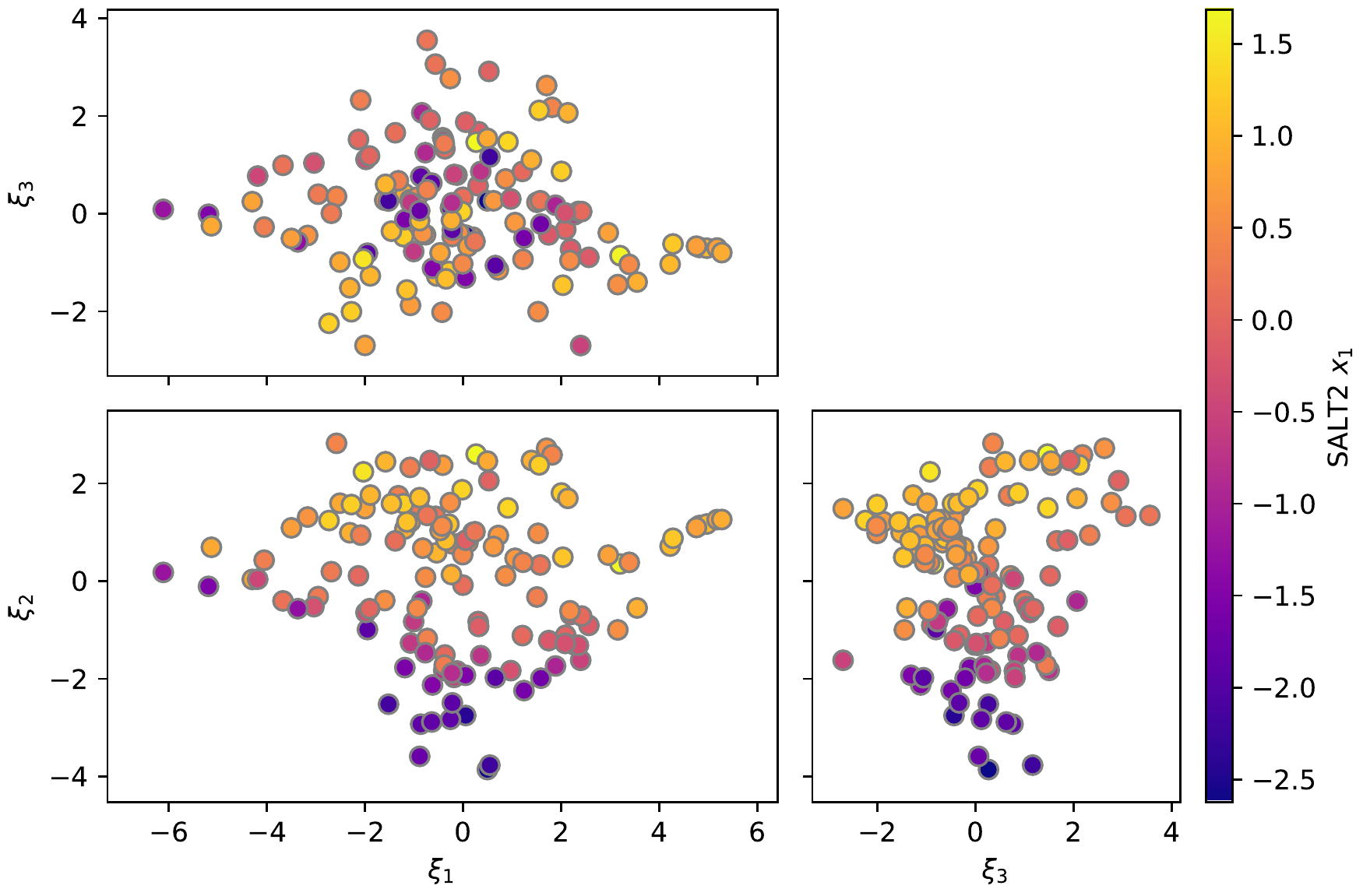}
    \caption{
        Values of SALT2 $x_1$ as a function of the Twins Embedding coordinates $\xi_i$. Each data point is a single SN~Ia,
        colored according to its measured SALT2 $x_1$ value.
    }
    \label{fig:salt2_x1_components}
\end{figure*}

For comparison purposes, we show the distribution of the SALT2 $x_1$ parameter over the Twins Embedding in
Figure~\ref{fig:salt2_x1_components}. Note that $x_1$ is highly correlated with $\xi_2$, so traditional SALT2
standardization will effectively include a linear correction in $\xi_2$. For low values of $x_1$, the implied GP
magnitude residuals are similar for all SNe~Ia with similar values of $x_1$. For high values of $x_1$, however,
the implied GP magnitude residuals vary fairly significantly as a function of the first Twins Embedding
component ($\xi_1$) in
a direction that is orthogonal to the variation in $x_1$. This will be discussed in more detail in
Section~\ref{sec:salt_biases}.

We investigated whether there is any improvement to the Twins Embedding standardization
if we include a correction for the SALT2 $x_1$ parameter in the model. With a linear correction in
SALT2 $x_1$, we find that the RMS is improved by only 0.001~mag for the RBTL + Twins Embedding model
and 0.004~mag for the SALT2 + Twins Embedding model. This negligible improvement can be explained
by the fact that $x_1$ is highly correlated with $\xi_2$, so including $x_1$ in the standardization
model does not add any new information.

\section{Discussion} \label{sec:discussion}

\subsection{Comparison of Standardization Techniques} \label{sec:standardization_comparison}

In Section~\ref{sec:standardization}, we established several different standardization techniques. In this Section,
we compare the magnitude residuals from these different techniques. We label the standardization techniques
from each section with short names as shown in Table~\ref{tab:standardization_labels}.

\begin{deluxetable}{ll}
\tablecaption{
    Labels for the different standardization methods described in Section~\ref{sec:standardization}
}
\label{tab:standardization_labels}
\tablehead{
    \label{Standardization Method Name} & \label{Section}
}

\startdata
    SALT2 + $x_1$ & Section~\ref{sec:salt2_standardization} \\
    Raw RBTL & Section~\ref{sec:raw_rbtl} \\
    RBTL + Twins Embedding & Section~\ref{sec:rbtl_standardization} \\
    SALT2 + Twins Embedding & Section~\ref{sec:salt2_isomap_standardization} \\
\enddata

\end{deluxetable}

We show the measured unweighted RMS and NMAD of the corrected magnitude residuals for all of these techniques in
Table~\ref{tab:standardization_comparison}. These measured RMS values have a contribution
from the peculiar velocities of the host galaxies of the SNe~Ia, as described in
Section~\ref{sec:magnitude_requirements}. Assuming a 300~km/s dispersion in velocity, and taking
the redshift of each SN~Ia in the sample into account, this contributes an
added dispersion of \pecvelcontribution~mag to the quoted RMS for the full sample. This
dispersion can be removed in quadrature from the
quoted values to obtain an estimate of the RMS of magnitude residuals that would be obtained for
samples of SNe~Ia at higher redshifts where peculiar velocity uncertainties have less of an impact on the
magnitude residuals. We also show these ``peculiar velocity removed'' RMS values in
Table~\ref{tab:standardization_comparison}.

\begin{deluxetable*}{llrcccc}
\tablecaption{
    Comparison of standardization performance for the different standardization methods
    and selection requirements. For each method, we show the NMAD, RMS, and unexplained
    dispersion of the corrected magnitude residuals along with an
    estimate of the RMS with the peculiar velocity removed (see text for details).
    Note that with the same sets of cuts, the quoted uncertainties are
    highly correlated between the different standardization techniques, and therefore do not apply to
    relative comparisons between different techniques.
}
\label{tab:standardization_comparison}
\tablehead{
    \colhead{Selection}    & \colhead{Number of} & \colhead{Statistic}    & \colhead{Raw RBTL}         & \colhead{RBTL + Twins}            & \colhead{SALT2 + $x_1$}     & \colhead{SALT2 + Twins}          \\[-0.5em]
    \colhead{Requirements} & \colhead{SNe~Ia}    &                        & \colhead{Dispersion}       & \colhead{Embedding}    &  \colhead{Dispersion}      & \colhead{Embedding}\\[-0.5em]
                           &                     &                        &  & \colhead{Dispersion}  & & \colhead{Dispersion} \\[-0.5em]
                           &                     &                        & \colhead{(mag)} & \colhead{(mag)}  & \colhead{(mag)}  & \colhead{(mag)}
}

\startdata
                  Full sample &   134 &                 NMAD &   0.108 $\pm$  0.013 &  0.083 $\pm$  0.010 &             \nodata &             \nodata \\
                          &       &                  RMS &   0.131 $\pm$  0.010 &  0.101 $\pm$  0.007 &             \nodata &             \nodata \\
                          &       & Peculiar velocity removed &   0.119 $\pm$  0.011 &  0.084 $\pm$  0.009 &             \nodata &             \nodata \\
 & & Unexplained dispersion & \nodata &   0.073 $\pm$  0.008 & \nodata & \nodata \\
\hline
                   Sample &   127 &                 NMAD &   0.111 $\pm$  0.012 &  0.084 $\pm$  0.011 &  0.106 $\pm$  0.012 &  0.105 $\pm$  0.013 \\
               with valid &       &                  RMS &   0.131 $\pm$  0.011 &  0.100 $\pm$  0.008 &  0.140 $\pm$  0.012 &  0.118 $\pm$  0.008 \\
               SALT2 fits &       & Peculiar velocity removed &   0.120 $\pm$  0.012 &  0.084 $\pm$  0.009 &  0.129 $\pm$  0.013 &  0.104 $\pm$  0.009 \\
 & & Unexplained dispersion & \nodata &   0.072 $\pm$  0.008 &   0.118 $\pm$  0.015 &   0.085 $\pm$  0.010 \\
\hline
          Training sample &    72 &                 NMAD &   0.100 $\pm$  0.019 &  0.072 $\pm$  0.013 &             \nodata &             \nodata \\
                          &       &                  RMS &   0.126 $\pm$  0.013 &  0.099 $\pm$  0.011 &             \nodata &             \nodata \\
                          &       & Peculiar velocity removed &   0.112 $\pm$  0.015 &  0.081 $\pm$  0.014 &             \nodata &             \nodata \\
 & & Unexplained dispersion & \nodata &   0.073 $\pm$  0.012 & \nodata & \nodata \\
\hline
        Validation sample &    62 &                 NMAD &   0.113 $\pm$  0.017 &  0.086 $\pm$  0.014 &             \nodata &             \nodata \\
                          &       &                  RMS &   0.137 $\pm$  0.016 &  0.102 $\pm$  0.012 &             \nodata &             \nodata \\
                          &       & Peculiar velocity removed &   0.127 $\pm$  0.017 &  0.088 $\pm$  0.014 &             \nodata &             \nodata \\
 & & Unexplained dispersion & \nodata &   0.069 $\pm$  0.012 & \nodata & \nodata \\
\hline

\enddata

\end{deluxetable*}

We find that RBTL + Twins Embedding standardization
significantly outperforms both of the SALT2 standardization methods. The variance of the magnitude residuals from
the SALT2 + $x_1$ standardization is $\sim 2.4$ times that from RBTL + Twins Embedding standardization, meaning
that a SN~Ia
corrected with RBTL + Twins Embedding standardization will have $\sim 2.4$ times the weight in a cosmological analysis compared to
a supernova corrected with SALT2 + $x_1$ standardization. Note that the quoted uncertainties on the measured 
RMS do not apply for comparisons between different standardization methods because we are examining
the same set of supernovae and most of the residual variation is highly correlated. The RMS for RBTL + Twins Embedding
standardization is \saltrbtlrmsdiff~mag lower than the RMS for SALT2 + $x_1$ standardization, which is an
improvement with a significance of $\saltrbtlrmsdiffsig\sigma$.

One somewhat unexpected result is that we find better dispersions for the Raw RBTL magnitude residuals
compared to SALT2 + $x_1$ corrected magnitudes (RMS of \rawrbtlmagstdsaltcuts~mag compared to
\saltparamrms~mag for the set of SNe~Ia with valid SALT2 fits). This suggests that even without applying
any additional corrections using the intrinsic diversity of the SNe~Ia, the base RBTL method standardizes
SNe~Ia better than does SALT2 with corrections for intrinsic diversity. This was
also seen in \citetalias{fakhouri15}, where even the worst twins had a smaller dispersion in brightness than
the SALT2 fits to that same dataset. Some of this difference can be explained by the fact that the RBTL
algorithm was constructed to identify the regions of the spectrum of SNe~Ia with low intrinsic diversity, and then
use these regions of the spectrum to estimate the peak brightness and extinction. Hence its estimate of the peak
brightness is less affected by intrinsic diversity than an estimate from the light curve itself.
However, we also see that the RBTL + Twins Embedding RMS of \rbtlgprmssaltcut~mag is significantly lower
than the SALT2 + Twins Embedding RMS of \saltgprms~mag for the same sample of SNe~Ia. This suggests that the RBTL
algorithm produces estimates of the brightnesses and extinctions of SNe~Ia that are more robust than the ones from a
light curve fitter like SALT2.

These results can be compared to the analysis of \citet{bailey09} who showed that the ratio of fluxes
at 6420~\AA\ and 4430~\AA\ can standardize SNe~Ia to within $0.125 \pm 0.011$~mag. The RBTL algorithm can be thought of as an
extension of this procedure where, instead of identifying a single pair of wavelengths where there is low
intrinsic diversity, all such wavelengths are used.

\subsection{Biases in SALT2 Standardization} \label{sec:salt_biases}

\begin{figure*}
\epsscale{0.8}
\plotone{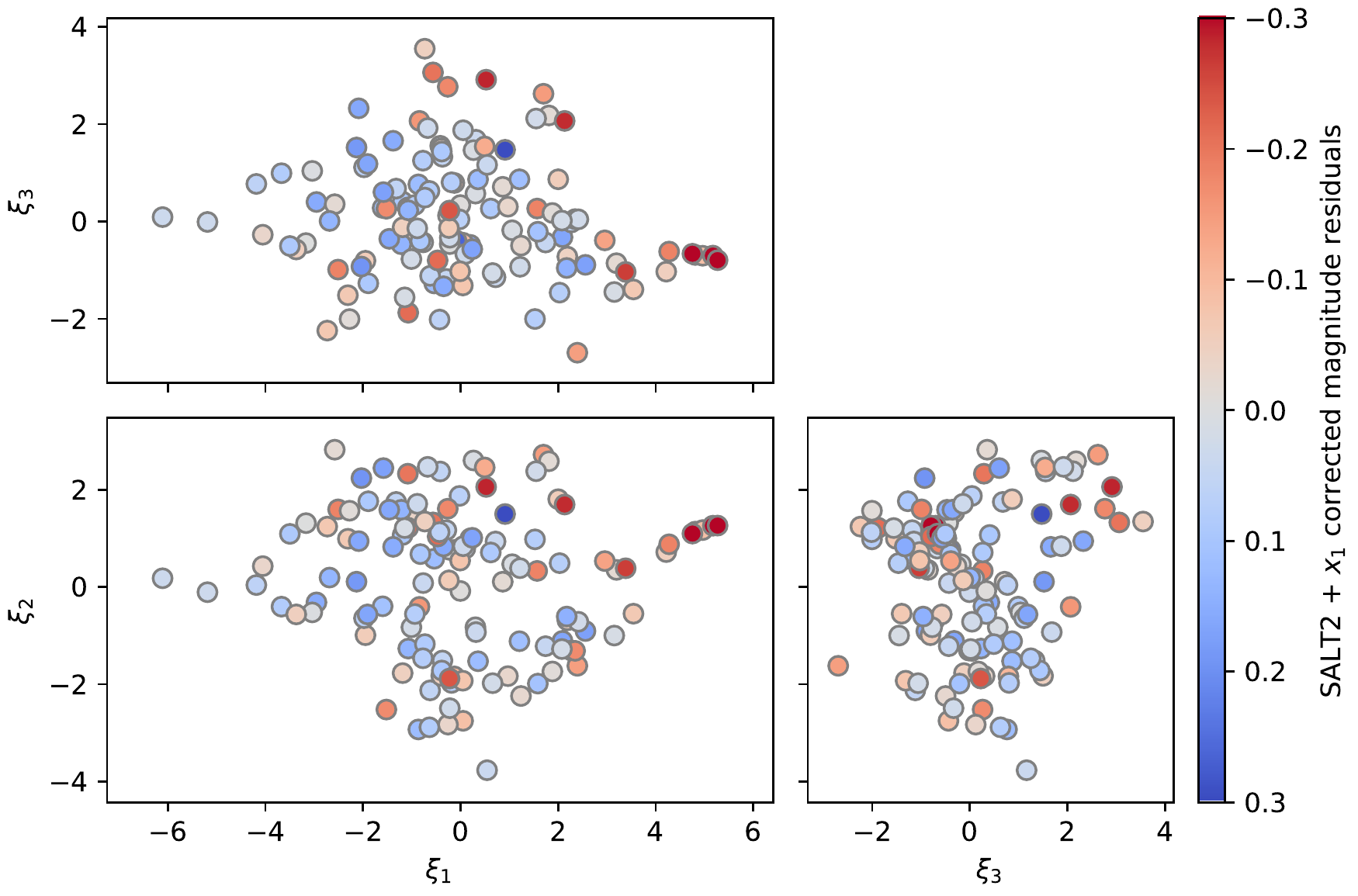}
\caption{
    SALT2 + $x_1$ corrected magnitude residuals as a function of the Twins Embedding
    coordinates $\xi_i$. We see noticeable structure in the residuals, especially for large
    values of the first and third Twins Embedding components ($\xi_1$ and $\xi_3$).
}
\label{fig:salt_residuals_manifold}
\end{figure*}

We can use the Twins Embedding to look for evidence of biases in SALT2 + $x_1$ standardization.
We show the SALT2 + $x_1$ magnitude residuals as a function of the Twins Embedding components in
Figure~\ref{fig:salt_residuals_manifold}. We see noticeable structure in the residuals,
especially for large values of the first and third Twins Embedding components ($\xi_1$ and $\xi_3$).

\begin{figure}
\plotone{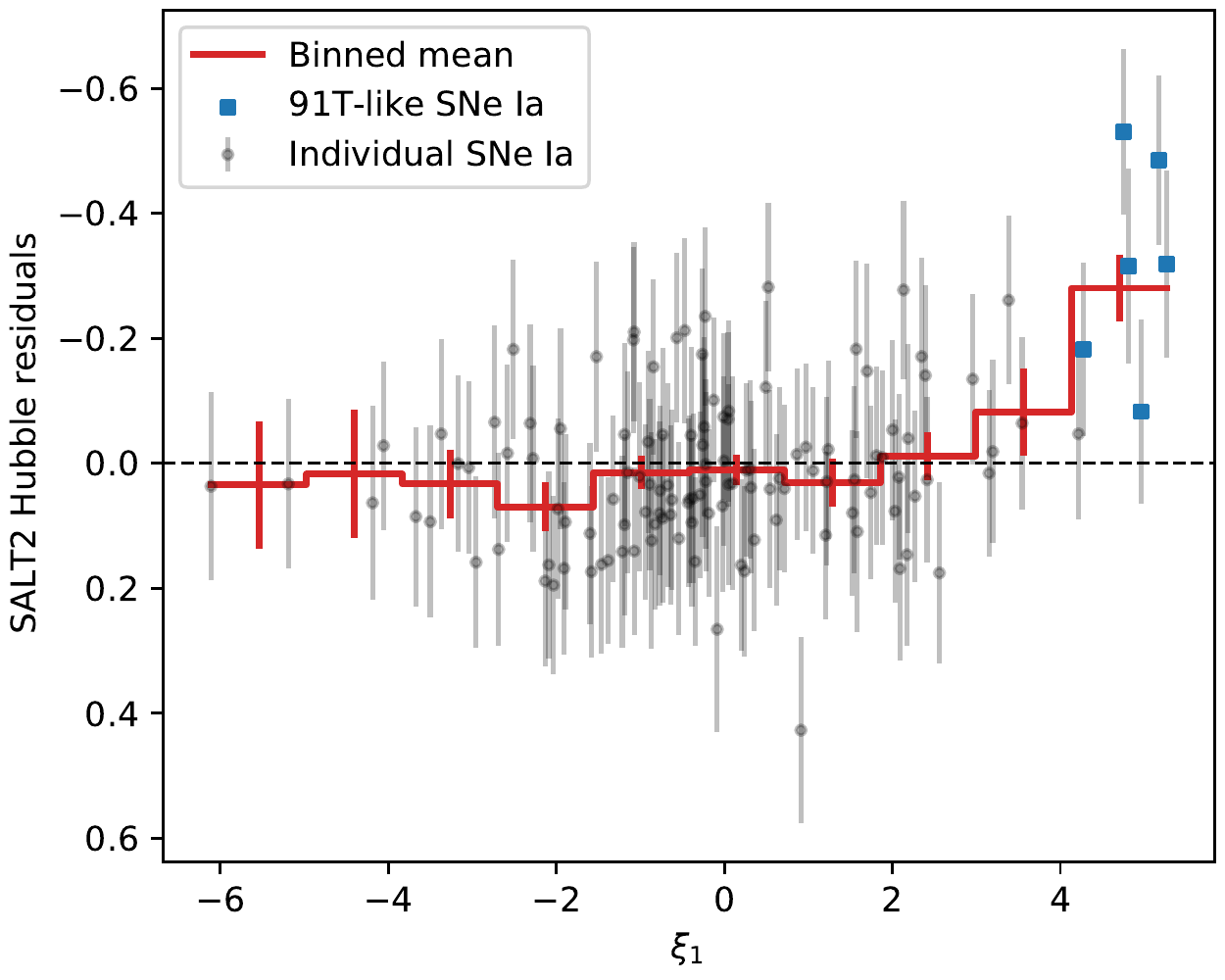}
\caption{
    SALT2 + $x_1$ corrected magnitude residuals as a function of the first component of
    the Twins Embedding $\xi_1$. The SALT2 magnitude residuals are shown for individual supernovae
    as gray points with their associated uncertainties. We calculate the mean magnitude
    residual and its uncertainty in evenly-spaced bins of this component. The results of that 
    procedures are shown with red lines. We find that SALT2 magnitude residuals show a strong bias
    for large values of $\xi_1$. Extreme values of this component are 91T-like SNe~Ia, as labeled
    with blue squares on the plot.
}
\label{fig:salt_hr_component_1}
\end{figure}

To probe the significance of this result, we examine the SALT2 + $x_1$ magnitude residuals
with their associated uncertainties as a function of the first Twins Embedding component ($\xi_1$) in
Figure~\ref{fig:salt_hr_component_1}. 
We label the 91T-like SNe~Ia in our sample in this Figure using the labels from
\citet{lin20submitted}. Extreme values of $\xi_1$ correspond to 91T-like SNe~Ia.
However, we see evidence of a bias even for SNe~Ia located at
$\xi_1$ values between the 91T-like SNe~Ia and the rest of the sample. When comparing SNe~Ia with
$\xi_1 > 3$ to SNe~Ia with $\xi_1 < 3$,
we find an offset in SALT2 + $x_1$ magnitude residuals of \saltfirstcompdiff\ mag. Note that the
location of this cut was selected after examining the distribution, which reduces the statistical power
of this measurement. However, this strongly suggests that the Twins Embedding
is identifying intrinsic diversity of SNe~Ia that affects the peak brightness estimates of SNe~Ia
and that is not captured by measures of the light curve width such as SALT2 $x_1$.

Cosmological analyses using SALT2 for standardization will not be able to identify
this subset of SNe~Ia as they have normal values of SALT2 $x_1$, as shown in
Figure~\ref{fig:salt2_x1_components}. This could lead to large systematic biases in cosmological analyses if
the relative abundance of this subset changes with redshift. 91T-like SNe~Ia are often associated
with active star formation \citep{hakobyan20}, which increases with redshift.
So, for example, if SNe~Ia in this subset were twice as
abundant at high redshifts compared to low redshifts, then there would be a systematic bias of $\sim$1\%
in the cosmological distance measurements. This is larger than the projected uncertainties from
upcoming surveys with the Vera C. Rubin Observatory \citep{lsst09} or the Nancy Grace Roman Space
Telescope \citep{hounsell18}. These biases could also exist for surveys that target specific kinds of
galaxies, such as Cepheid-hosting galaxies for measurements of the Hubble constant, if the
fraction of SNe~Ia belonging to the biased subset differs for SNe~Ia in these galaxies compared to
the overall sample of SNe~Ia.

As shown in Article I, there are several other indicators of intrinsic diversity that are highly
correlated with $\xi_1$ and that could be used to identify this subset of SNe~Ia. Notably, \citet{nordin18}
showed that their ``uCa'' feature (the flux of a SN~Ia in the U-band between 3750 and 3860~\AA) can be used
to identify 91T-like SNe~Ia and improve standardization. The uCa feature is correlated with $\xi_1$ and
performs a similar role when used for standardization.
Other features that could be used to identify these SNe~Ia include the pseudo-equivalent width of the
\ion{Ca}{2} H\&K feature, the pseudo-equivalent width of the \ion{Si}{2} 6355~\AA\ feature, the second
component ($q_2$) of the SUGAR model \citep{leget20}, or the first component ($c_1$) of the SNEMO7
model \citep{saunders18}.

\begin{figure}
\epsscale{1.15}
\plotone{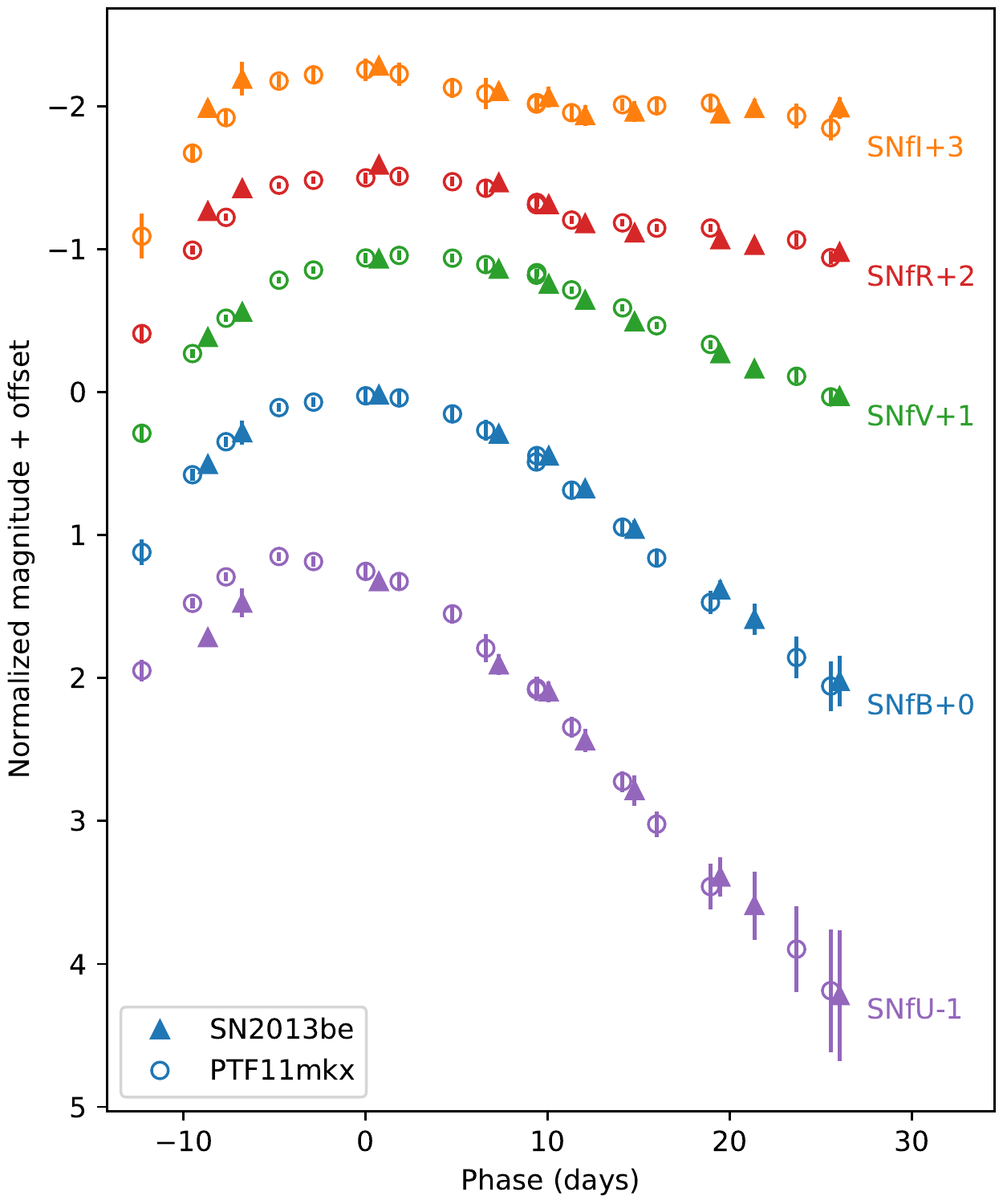}
\caption{
    Comparison of the photometry of \saltcompnamea\ and \saltcompnameb. We 
    show restframe synthetic photometry in the SNfactory bands defined in Table~\ref{tab:snfactory_bands}.
    The light curves have been normalized to their individual peak brightnesses in the \Bsnf band. The
    light curves are nearly identical in the \Bsnf and \Vsnf bands (and best-fit SALT2 parameters), but we
    see significant differences in the other bands.
}
\label{fig:salt_comparison}
\end{figure}

These results imply that a SALT2-like model with one component for color and another component
for the light curve width is not sufficient to fully capture the intrinsic diversity of SNe~Ia relevant
to standardization. To emphasize this point, we compare the normal SN~Ia \saltcompnamea\ to the 91T-like
SN~Ia \saltcompnameb. These
two SNe~Ia have nearly identical SALT2 parameters, with $x_1$ values of \saltcompxonea\ and
\saltcompxoneb, and $c$ values of \saltcompca\ and \saltcompcb\ respectively. We show the light
curves of these two SNe~Ia in the SNfactory bands (defined in Table~\ref{tab:snfactory_bands}) in
Figure~\ref{fig:salt_comparison}. The \Bsnf and \Vsnf-band light curves of these two SNe~Ia are nearly identical,
with differences of less than 0.1~mag for the first 30 days of the light curve. Measurements of
the light curve width or other properties in similar bands cannot distinguish between these two SNe~Ia.
These two SNe~Ia do still have significant differences in their light curves: the \Usnf, \Rsnf, and
\Isnf-band light curves differ by $\sim$0.3~mag before maximum light, and there are differences in the
profile of the secondary maximum in the redder bands, which
SALT2 does not model and which are not generally measurable for high-redshift SNe~Ia.

\begin{figure*}
\plotone{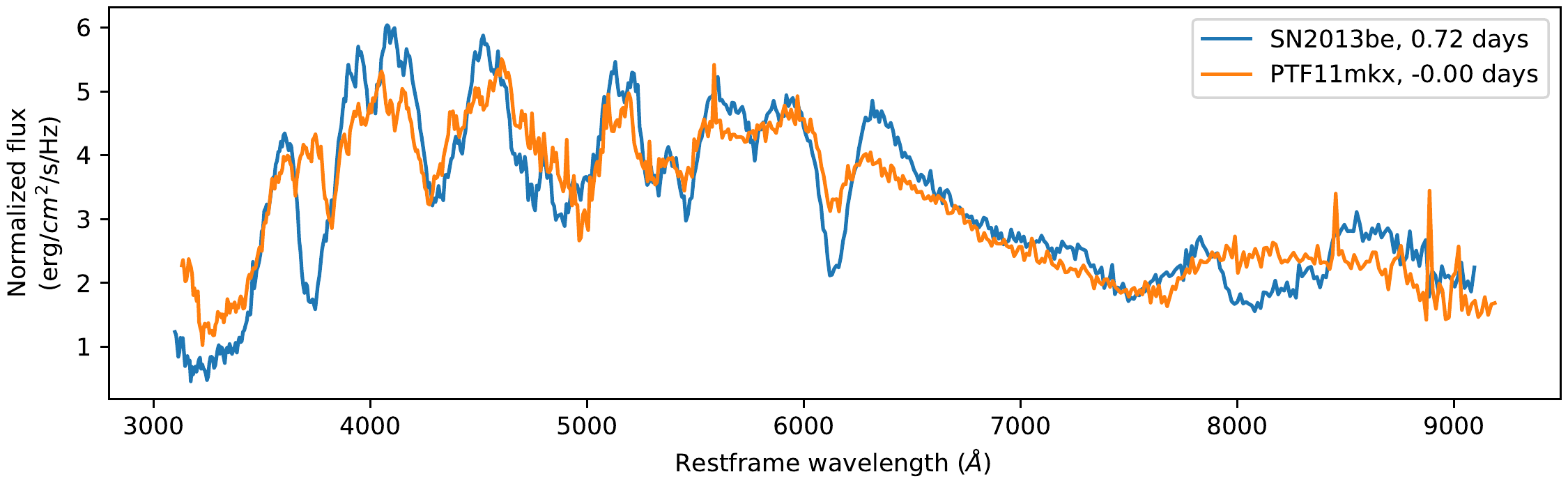}
\caption{
    Comparison of the spectra of \saltcompnamea\ and \saltcompnameb\ closest to maximum light.
    These spectra are shifted to the restframe, scaled so that they have the same brightness in the
    \Bsnf band, and binned to 500~km/s for visual purposes. There are large differences between the spectra
    in almost all of the spectral features, but they have nearly identical SALT2 $x_1$ and $c$
    parameters.
}
\label{fig:same_salt_spectra}
\end{figure*}

The SALT2 fits for the peak \Bsnf-band luminosities of these two light curves
differ by \saltcompmagdiff~mag, with the quoted uncertainty including both measurement uncertainties and potential
contributions from peculiar velocities. Because they have nearly identical $x_1$ and $c$ values, this
difference in the luminosities cannot be identified or corrected using SALT2. 
We show the restframe spectra of these two SNe~Ia closest to maximum light in
Figure~\ref{fig:same_salt_spectra}. There are large differences in nearly every single spectral feature.
This is captured by the Twins Embedding: these two SNe~Ia have $\xi_1$ values of \saltcompcoorda\ and
\saltcompcoordb\ respectively. With SALT2 + Twins Embedding standardization, as described in
Section~\ref{sec:salt2_isomap_standardization}, we find a difference between the corrected magnitude
residuals of these two SNe~Ia of \saltcompmanifoldmagdiff~mag. Hence standardization using the Twins Embedding
is able to correctly identify the difference in the luminosities of these two SNe~Ia while SALT2-like
standardization is not.

\subsection{Sufficiency of the Twins Embedding}

A similar question is whether the Twins Embedding is sufficient for
describing the diversity of spectra at maximum light and whether this impacts standardization.
The Twins Embedding was
constructed to preserve the spectral distances of \citetalias{fakhouri15}, and the Euclidean
distance between two SNe~Ia in the Twins Embedding would be equal to their spectral distance
if we kept all of the components of the embedding. We have, however, restricted ourselves
to a three component embedding so there could be some SNe~Ia that are not well-described
by our low-dimensional model.

\begin{figure*}[h]
\epsscale{1.1}
\plotone{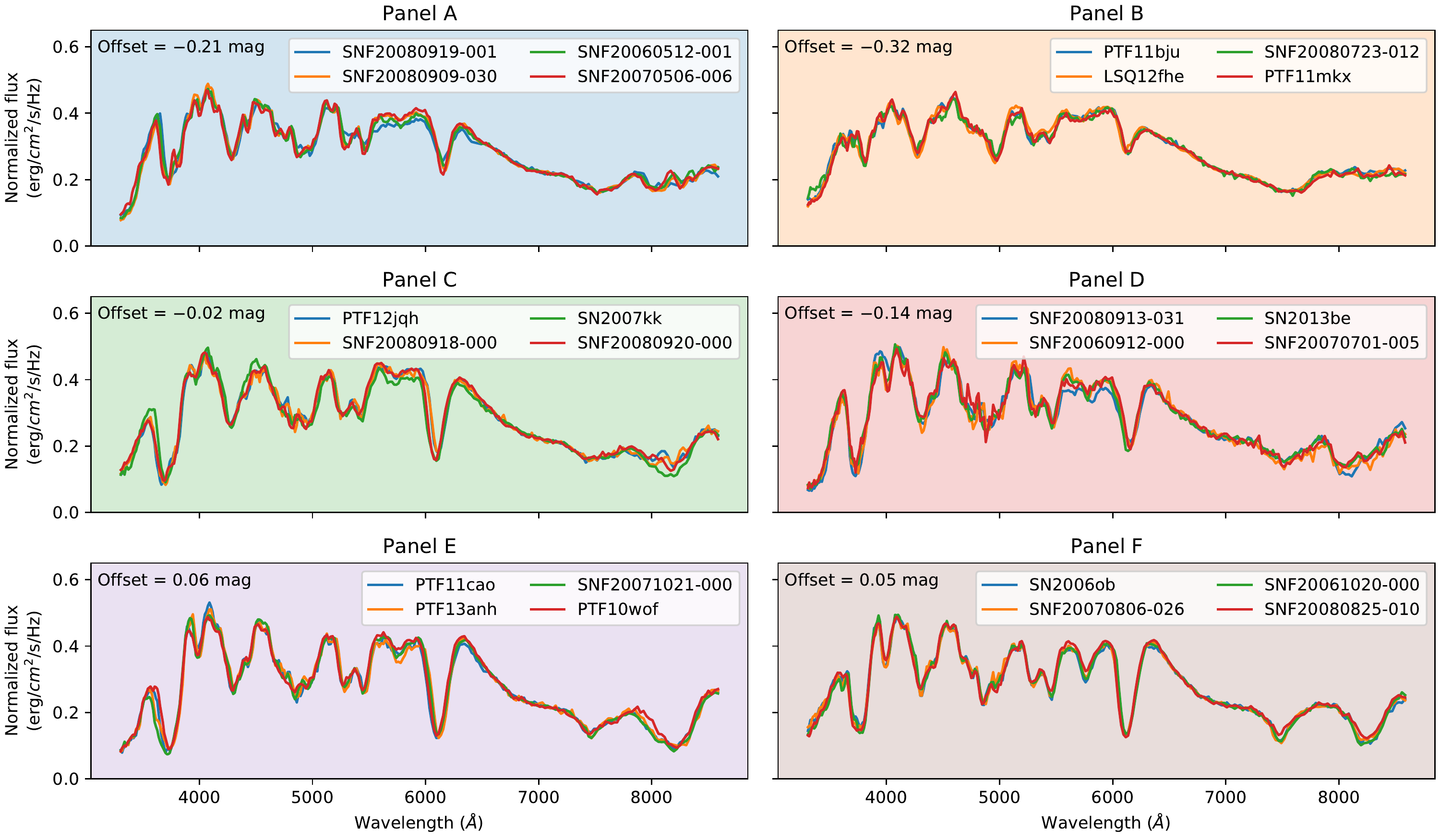} \\
\vspace{0.5cm}
\epsscale{0.8}
\plotone{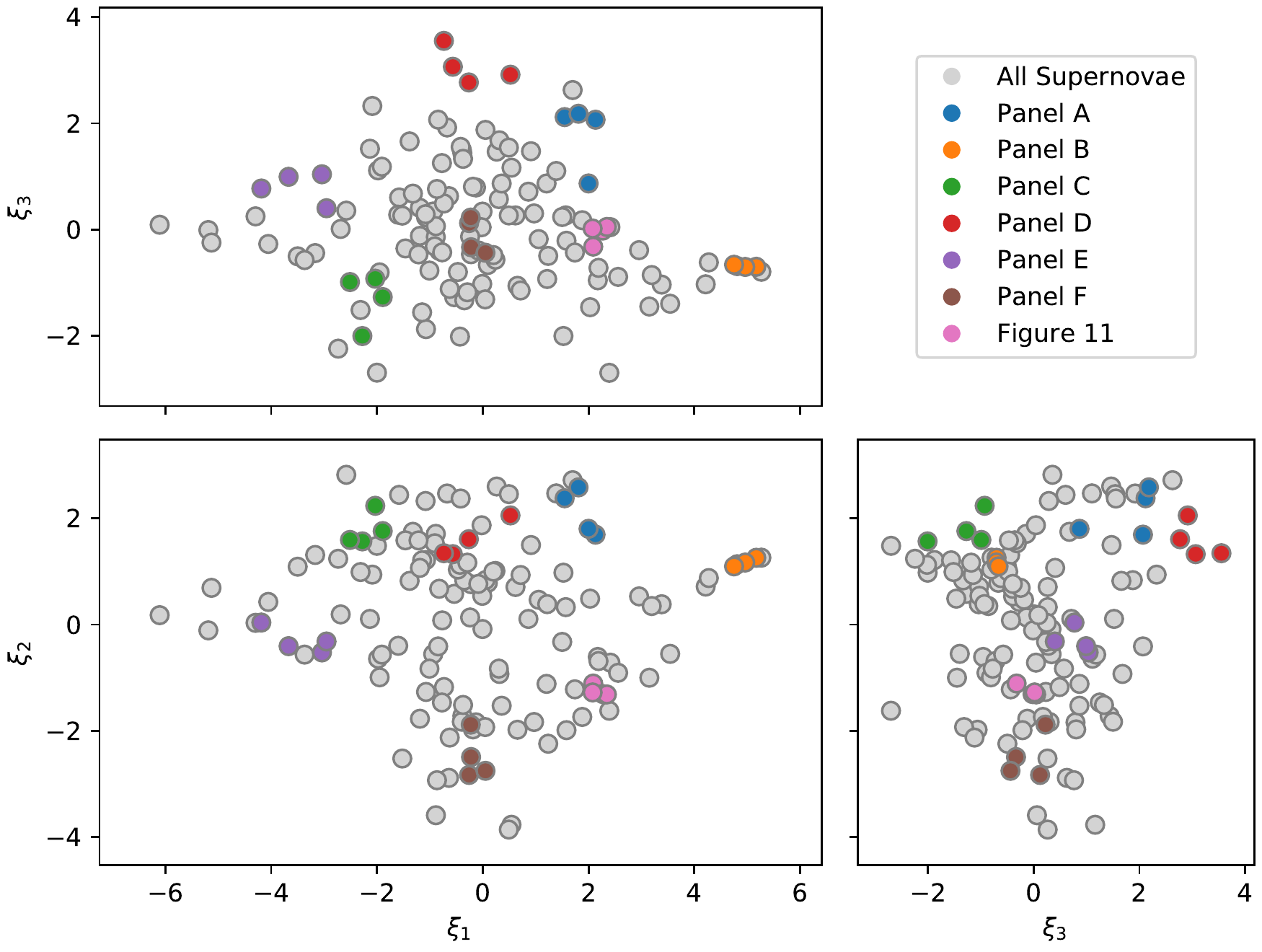}
\caption{
    Examples of the spectra of groups of SNe~Ia with similar Twins Embedding coordinates.
    In the bottom plot, we show the locations of each group in different colors. In the top
    plot we show the estimated spectra at maximum light of four SNe~Ia from each group binned
    to 1000~km/s and with the RBTL brightness and color removed. n the top left corner
    of each panel, we show the estimated offset in brightness for SNe~Ia in each group from RBTL + Twins Embedding
    standardization. The colors of the panels in
    the top plot correspond to the markers with the same colors in the bottom plot. We find
    that the spectra are remarkably similar within each of the groups.
}
\label{fig:spectra_groups}
\end{figure*}

We show the estimated spectra at maximum light for six different groups of
SNe~Ia that are nearby in the Twins Embedding in Figure~\ref{fig:spectra_groups}. For
each of the groups, the spectra of all of the different SNe~Ia are remarkably similar with only
slight differences around major spectral features. The magnitude residuals for RBTL + Twins Embedding
standardization are consistent with zero for each group, in agreement with the discussion
in Section~\ref{sec:rbtl_standardization}.

\begin{figure*}
\plotone{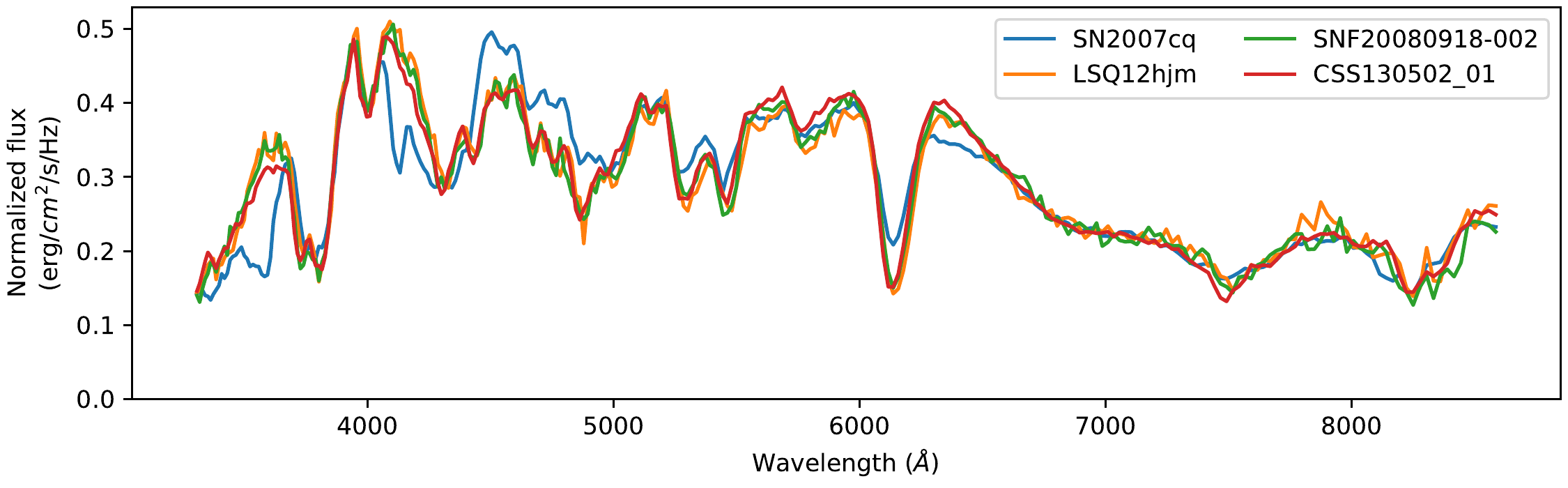}
\caption{
    Comparison of the spectra of SNe~Ia that are closest to SN2007cq in the Twins Embedding.
    The spectra are binned to 1000~km/s and the RBTL brightness and color have been removed.
    These spectra have nearly identical Twins Embedding coordinates and SALT2 $x_1$ parameters,
    but SN2007cq shows absorption features around 3600 and 4130~\AA\ that we identify as
    \ion{Ti}{2} that are not seen in the other spectra. As discussed in the test, SN2007cq
    is the largest spectral outlier for this analysis, but it has a normal magnitude residual
    and is not an outlier for RBTL + Twins Embedding standardization.
}
\label{fig:sn2007cq_neighbors}
\end{figure*}

We looked for examples of pairs of SNe~Ia that were in the worst 50\% of \citetalias{fakhouri15}
spectral distances, but in the best 10\% of distances in the Twins Embedding. We found nine such
pairings out of the 8,001 pairings considered, all but one of which include SN2007cq. We show the spectrum
of SN2007cq and three other nearby SNe~Ia that are nearby in the Twins Embedding in Figure~\ref{fig:sn2007cq_neighbors}.
The spectrum of SN2007cq shows absorption features near 3600 and 4130~\AA\ that
we identify as \ion{Ti}{2} that are similar to what is seen in the spectra of 91bg-like SNe~Ia. At redder
wavelengths, the spectrum of SN2007cq is similar to those of core normal SNe~Ia and it does not show the
deep \ion{Si}{2} and \ion{Ca}{2} absorption features that are
typical of 91bg-like SNe~Ia.

SN2007cq and its three closest neighbors have very similar SALT2 $x_1$ values, between
$-0.31$ and $-0.74$, and there are no major differences in their light curves. Despite its unusual
spectrum, SN2007cq is not a standardization outlier with RBTL + Twins Embedding standardization: it has
an relatively normal magnitude residual of $-0.18 \pm 0.12$~mag. The explosions of SNe~Ia are very complex, and we expect that there
will be some rare SNe~Ia whose diversity is not captured by our low dimensional model. SN2007cq is the most
egregious such spectral ``outlier'' in our current analysis, but our standardization methods still perform
adequately well for it. We do not find any examples of SNe~Ia that have both large magnitude residuals and large spectral
differences relative to nearby SNe~Ia in the Twins Embedding.

\subsection{Correlations with Host Galaxy Properties} \label{sec:host_properties}

As described in Section~\ref{sec:standard_candles}, SALT2 standardized magnitude residuals have been shown to
have differences of $\sim$0.1~mag when comparing SNe~Ia from host galaxies with different properties.
We examined how standardization using the Twins Embedding affects these correlations with host galaxy properties.
We use the measurements of host galaxy properties from \citet{rigault18} (hereafter \citetalias{rigault18}), which
includes much of the same set of SNe~Ia used in our analysis. The authors of this analysis found that there is a step of
$0.163 \pm 0.029$~mag when comparing SALT2 residuals from younger environments to older ones, measured using the
``local specific star formation rate'' (LsSFR). They also found a step of $0.119 \pm 0.032$~mag for high-mass hosts
compared to low-mass hosts.

We calculated the size of the steps for each of our standardization methods following the same procedure
as in \citetalias{rigault18} as a function of both the host mass and the LsSFR.
All SNe~Ia are assigned a probability of being on each side of the step given a threshold value and
the measurements of the SN~Ia host galaxies. We modify the standardization procedures described in
Section~\ref{sec:standardization} to simultaneously fit for the size of the host step as part of the
standardization procedure. As in \citetalias{rigault18}, we reject peculiar SNe~Ia for this
analysis, with ``peculiar'' referring to all SNe~Ia that are identified as 91bg-like,
02cx-like or 91T-like in \citet{lin20submitted}.

With SALT2 + $x_1$ standardization, we find a host step of \hoststepfitsaltlssfr\ mag for LsSFR
and \hoststepfitsaltmass\ mag for the host mass. These results are consistent with the results found in
\citetalias{rigault18}, and differ because we have different selection criteria for our analysis. When
standardizing these same SNe~Ia using RBTL + Twins Embedding standardization, we find that the host steps
are significantly decreased: we find step sizes of \hoststepfitrbtllssfr\ mag for LsSFR and
\hoststepfitrbtlmass\ mag for the host mass.

\begin{figure*}
\epsscale{1.1}
\plottwo{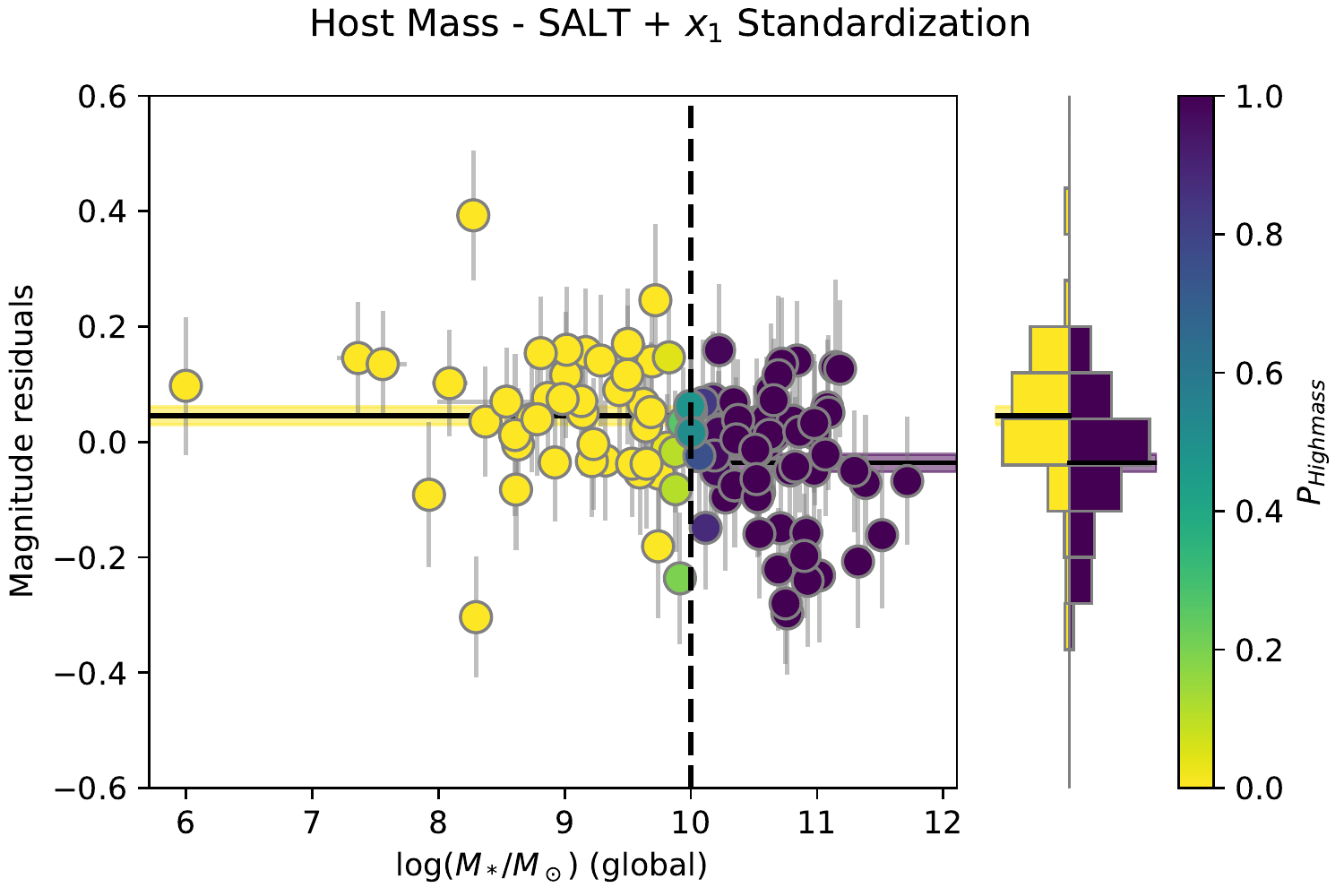}{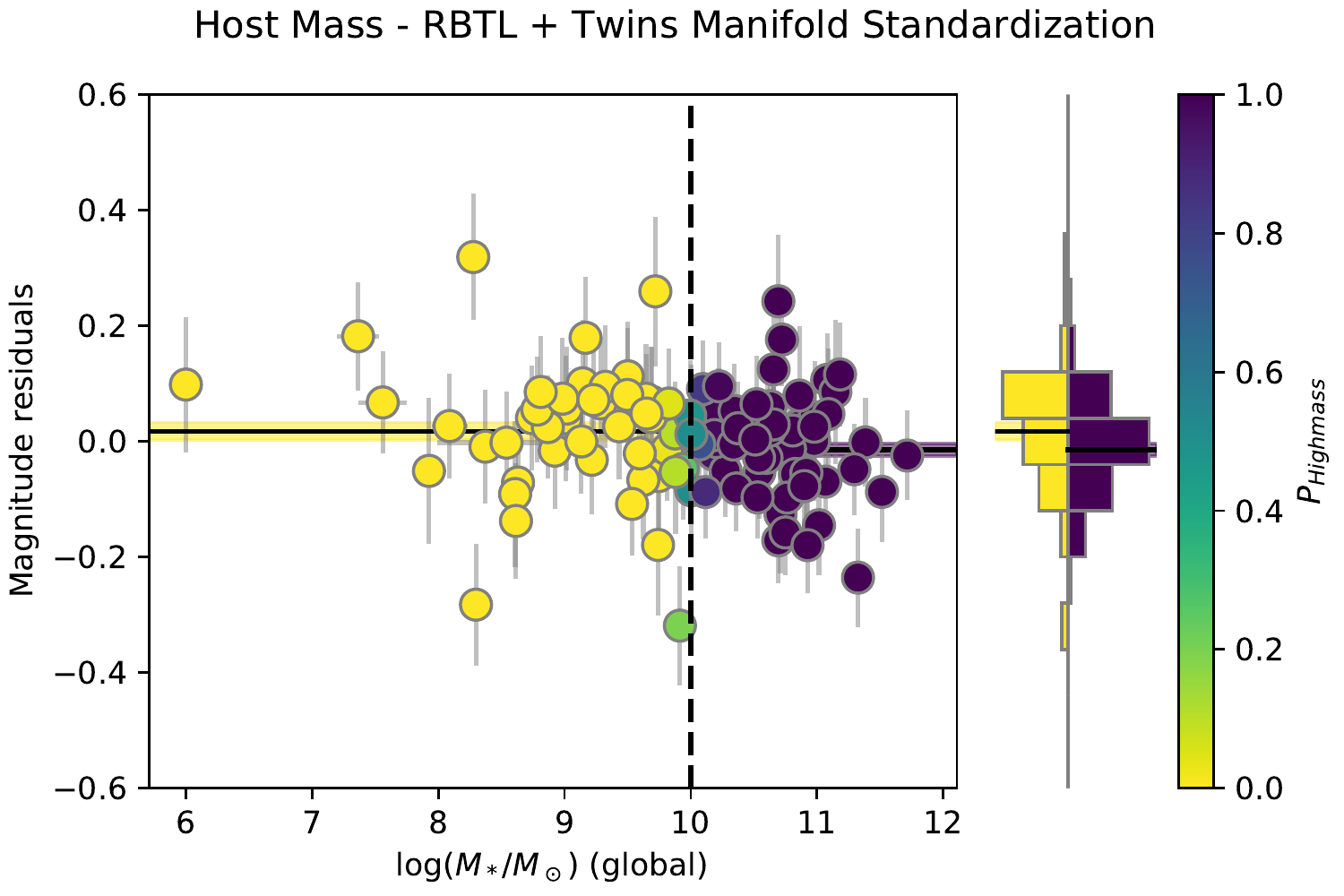} \\
\vspace{0.5cm}
\plottwo{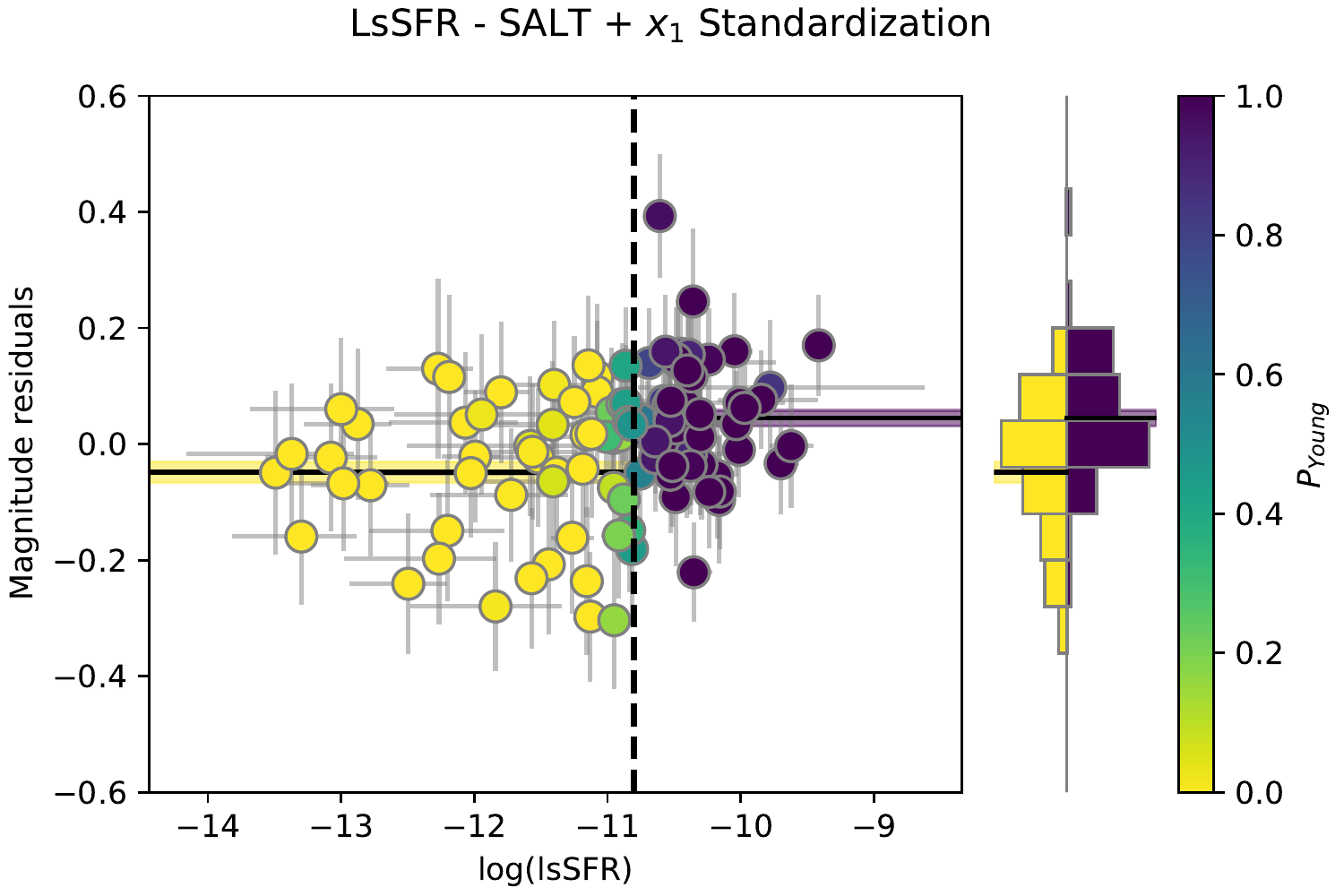}{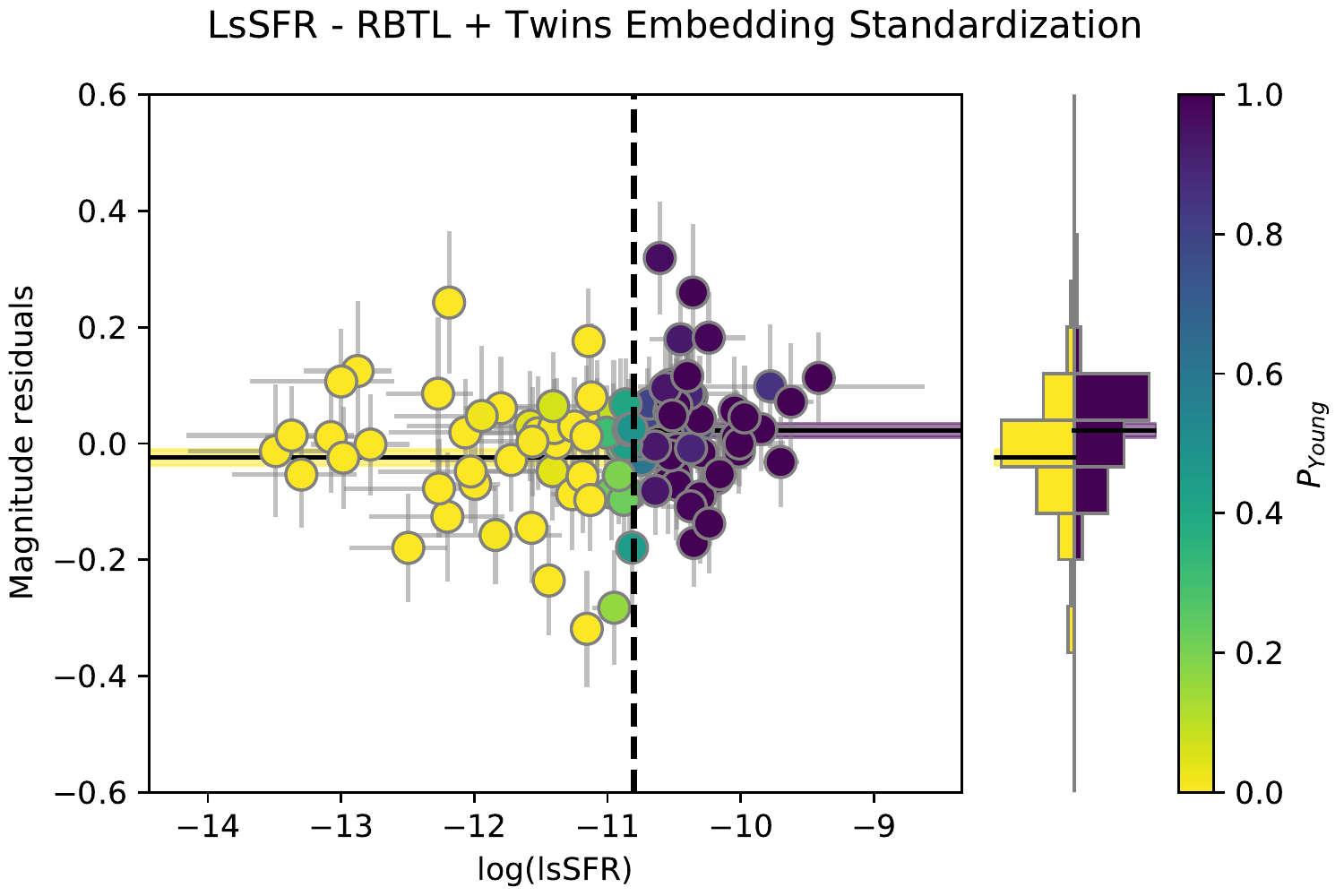}
\caption{
    Magnitude residuals as a function of host galaxy properties. The left two panels show
    SALT2 + $x_1$ corrected magnitude residuals while the right two panels show RBTL + Twins Embedding
    corrected magnitude residuals. The top two panels show the magnitude residuals as a function of the
    host galaxy mass while the bottom two panels show the magnitude residuals as a function of LsSFR.
    The marker color represents the probability of each SN~Ia being in a low/high mass galaxy or young/old
    region given the threshold shown as a vertical line. The histograms on the right side of each panel
    show the distribution of the magnitude residuals for SNe~Ia on either side of the threshold.
    The two horizonal bands show the measured mean magnitude residual for each group and its uncertainty,
    determined with a Gaussian mixture model as described in the text.
}
\label{fig:host_steps}
\end{figure*}

We also perform several variants on this analysis. First, instead of fitting the step sizes
as part of the standardization procedure, we examine the size of them after the standardization
correction has already been applied. We do this by fitting a Gaussian mixture model to
the corrected magnitude residuals with different means and standard deviations for the
magnitude residuals of SNe~Ia on either side of the step. We find that the LsSFR step decreases
from \hoststepgmmsaltlssfr\ mag to \hoststepgmmrbtllssfr\ mag between SALT2 + $x_1$ standardization
and RBTL + Twins Embedding standardization, and the host mass step decreases from
\hoststepgmmsaltmass\ mag to \hoststepgmmrbtlmass\ mag. We show the magnitude residuals and
the results of this procedure in Figure~\ref{fig:host_steps}. Note that the uncertainties in
the step
measurements are highly correlated. For this analysis variant, we used bootstrap resampling to
estimate the significance of the decrease in step size and find that it is significant at the
$ \hoststepdifflssfrsignificance \sigma $ level for both of the host variables.

\citetalias{rigault18} removed all peculiar SNe~Ia from their main analysis. We examine
how the step sizes are impacted when they are included. Interestingly, the host mass step for
SALT2 + $x_1$ standardization decreases significantly when we include peculiar SNe~Ia
from \hoststepfitsaltmass\ mag to \hoststeppeculiarsaltmass\ mag. The LsSFR step shows a small
decrease from \hoststepfitsaltlssfr\ mag to \hoststeppeculiarsaltlssfr\ mag.
\citetalias{rigault18} found similar results. The step sizes for RBTL + Twins Embedding standardization change by
less than 0.01\ mag. This can be explained by the fact that the SALT2 + $x_1$ corrected magnitude
residuals of 91T-like peculiar SNe~Ia are biased by $\sim$\saltfirstcompdiff\ mag as shown in
Section~\ref{sec:salt_biases}. For our sample of SNe~Ia, the 91T-like SNe~Ia are preferentially
in low mass/high LsSFR hosts, which artificially decreases the apparent size of
the host steps. This result implies that for SALT2 + $x_1$ standardization, the size of measured
host steps will vary depending on the fraction of 91T-like SNe~Ia in the sample. RBTL + Twins Embedding
standardization correctly handles 91T-like SNe~Ia and is not affected by this.

\begin{figure*}
\plotone{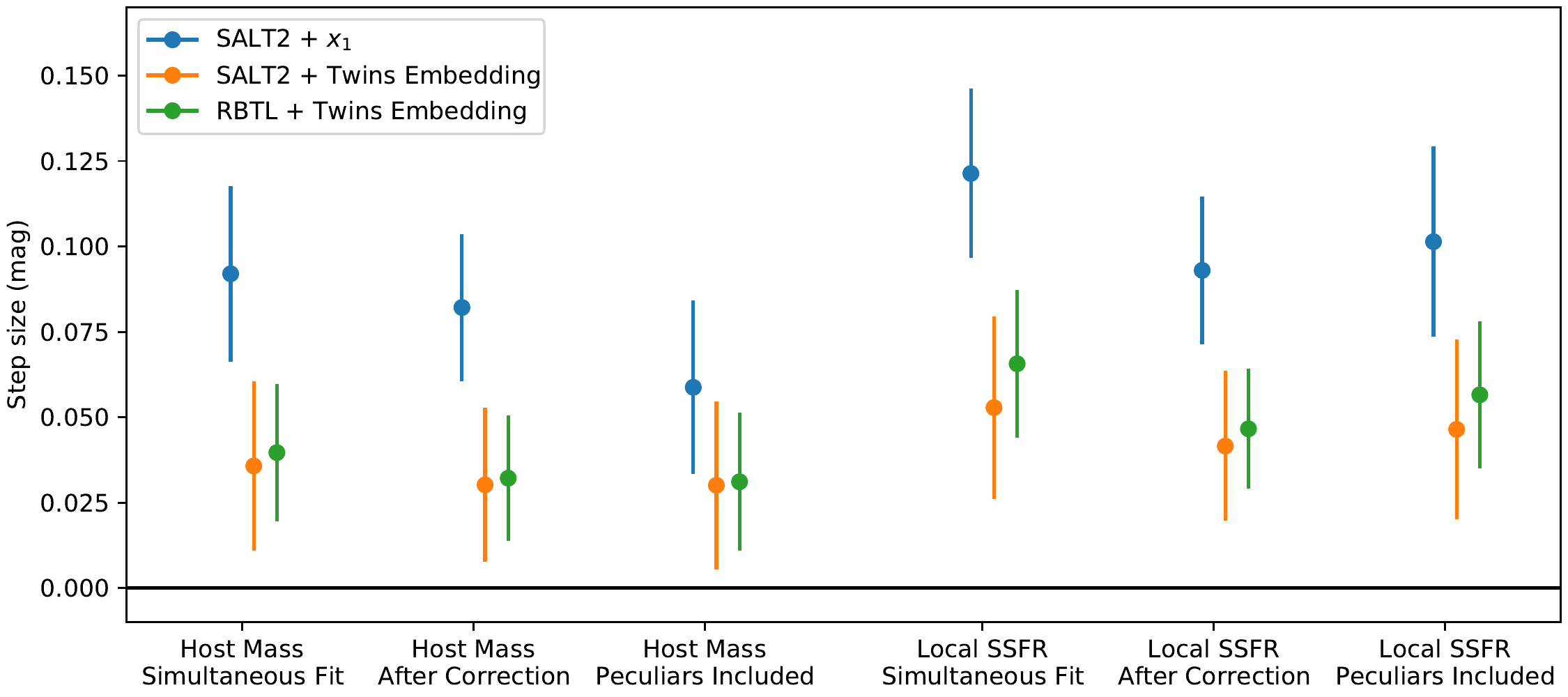}
\caption{
    Summary of the host step measurements for the different analysis variants discussed in
    Section~\ref{sec:host_properties}. For each analysis variant, we
    show the recovered step sizes for our three different standardization methods in
    different colors. Note that the uncertainties in the step sizes between different
    standardization methods are highly correlated: the decrease in host step size is
    significant at the 3.3--3.7$\sigma$ level for the measurements
    of host step sizes after correction.
}
\label{fig:host_step_comparison}
\end{figure*}

We measured the step sizes for SALT2 + Twins Embedding standardization, and find very
similar results to what is seen for RBTL + Twins Embedding standardzation. The measured host steps
for all of these analysis variants are shown in
Table~\ref{tab:host_step_comparison}, and a summary of the host step sizes is shown in
Figure~\ref{fig:host_step_comparison}.

\begin{deluxetable*}{llD@{ $\pm$}DD@{ $\pm$}DD@{ $\pm$}D}
\tablecaption{
    Measured host-galaxy property step sizes for different standardization methods. See text for details. We choose to set
    the sign of the step to be positive for the SALT2 + $x_1$ analysis: a negative step means that we recover a step in
    the opposite direction from the one for the SALT2 + $x_1$ analysis.
}
\label{tab:host_step_comparison}
\tablehead{
    \colhead{Analysis variant} & \colhead{Host property} & \multicolumn4c{SALT2 + $x_1$}  & \multicolumn4c{RBTL + Twins} & \multicolumn4c{SALT2 + Twins} \\[-0.5em]
    \colhead{} & & \multicolumn4c{step size (mag)} & \multicolumn4c{Embedding} & \multicolumn4c{Embedding} \\[-0.5em]
    \colhead{} & & \multicolumn4c{} & \multicolumn4c{step size (mag)} & \multicolumn4c{step size (mag)}
}
\decimals
\startdata
    Simultaneous fit & Host Mass & 0.092 & 0.024 & 0.036 & 0.025 & 0.040 & 0.020 \\
 & Local SSFR & 0.121 & 0.029 & 0.053 & 0.027 & 0.066 & 0.022 \\
\hline
After correction & Host Mass & 0.082 & 0.021 & 0.030 & 0.023 & 0.032 & 0.018 \\
 & Local SSFR & 0.093 & 0.022 & 0.042 & 0.022 & 0.047 & 0.018 \\
\hline
Peculiars included & Host Mass & 0.059 & 0.026 & 0.030 & 0.025 & 0.031 & 0.020 \\
 & Local SSFR & 0.101 & 0.028 & 0.046 & 0.026 & 0.057 & 0.022 \\
\hline
Training subset & Host Mass & 0.077 & 0.028 & -0.005 & 0.033 & 0.026 & 0.026 \\
 & Local SSFR & 0.124 & 0.037 & 0.050 & 0.037 & 0.056 & 0.029 \\
\hline
Validation subset & Host Mass & 0.110 & 0.048 & 0.064 & 0.035 & 0.058 & 0.029 \\
 & Local SSFR & 0.119 & 0.059 & 0.018 & 0.039 & 0.057 & 0.033 \\

\enddata
\end{deluxetable*}

\section{Conclusions} \label{sec:conclusions}

In this work, we introduced several methods that can be used to standardize SNe~Ia with significantly improved
performance compared to traditional SALT2 + $x_1$ standardization. With the ``Reading Between the Lines''
method, we can obtain a robust estimate of the peak brightness and extinction of a SN~Ia from a single
photometrically-calibrated spectrum at maximum light by using the regions of the spectrum with low intrinsic
diversity. We find that the RBTL method on its own is a very
good estimator of the distances to SNe~Ia, with a dispersion in the RBTL magnitude residuals of
\rawrbtlmagstd~mag. The RBTL algorithm can be applied to all SNe~Ia, including ones that are normally
labeled as ``peculiar''.

We showed that the Twins Embedding introduced in Article I can be used to take into account the intrinsic diversity and
standardize the distance estimates from either RBTL or SALT2. Using Gaussian process regression, we estimate
the magnitude residual for each SN~Ia from its local neighborhood in the Twins Embedding. This significantly
improves the standardization of these SNe~Ia: we find
an RMS dispersion in the corrected magnitude residuals of \saltparamrms~mag for conventional SALT2 + $x_1$ standardization
compared to \saltgprms~mag for SALT2 + Twins Embedding standardization, and \rbtlgprmssaltcut~mag for RBTL + Twins Embedding
standardization on the same set of SNe~Ia.

These dispersions all contain additional scatter due to host galaxy peculiar velocities.
For analyses of SNe~Ia at higher redshifts or
studies of the peculiar velocities themselves, our results imply that RBTL + Twins Embedding standardization
is accurate to within \rbtlgpcomppvremrms~mag compared to \saltcomppvremrms~mag for SALT2 + $x_1$ standardization.
Additionally, a significant fraction of this remaining dispersion is due to uncertainties in the GP model that
will be eliminated if this analysis is run using a larger sample of SNe~Ia. The remaining unexplained dispersion
is \rbtlgpintdisp~mag for RBTL + Twins Embedding standardization and \saltgpintdisp~mag for SALT2 + Twins Embedding
standardization compared to \saltparamsigmaint~mag for SALT2 + $x_1$ standardization. Note that the uncertainties
on all of these RMS dispersion values are correlated: the improved dispersion for RBTL + Twins Embedding
standardization relative to SALT2 + $x_1$ standardization is significant at the $\saltrbtlrmsdiffsig\sigma$ level.

This decreased dispersion implies that SNe~Ia standardized using RBTL + Twins Embedding standardization
have $\sim$2.4 times as much weight in a cosmology analysis compared to those standardized using SALT2 + $x_1$.
This is particularly of interest for studies of nearby SNe~Ia, such as measurements of the Hubble
constant, where the rate of SNe~Ia is limited but high-quality measurements are relatively inexpensive
to obtain.

This improved standardization is as important for the systematic uncertainties as it is for the 
statistical uncertainties, since the systematic uncertainties are constrained to fit within the
remaining \rbtlgpintdisp~mag of unexplained dispersion for the RBTL + Twins Embedding analysis. This is
a large improvement over the \saltparamsigmaint~mag of unexplained dispersion in the SALT2 + $x_1$ analysis.
This difference may be in part explained by our finding that SALT2 + $x_1$ standardization is biased by
\saltfirstcompdiff\ mag for a subset of
SNe~Ia that can be identified with the first component of the Twins Embedding $\xi_1$. This subset of SNe~Ia
includes, but is not limited to, 91T-like SNe~Ia. A major concern for cosmological analyses is that
the rates of SNe~Ia in different regions of the parameter space could evolve with redshift.
If SNe~Ia in this region of parameter
space have a different rate at high redshifts compared to low redshifts, then they could significantly bias
cosmological measurements. With a SALT2 light curve fit, these SNe~Ia are indistinguishable from some of
the ``normal'' SNe~Ia. However, we showed in this analysis that
we can distinguish them from the rest of the sample using the Twins Embedding that was constructed from
spectrophotometrically-calibrated spectra at maximum light. Future work needs to be done to determine
whether these biased subpopulations can be identified with other techniques such as
lower-resolution and lower-signal-to-noise spectroscopy or more advanced light curve fitters.

Upcoming surveys such as the Rubin Observatory's LSST will not have
spectrophotometrically-calibrated spectra near maximum
light for the vast majority of the SNe~Ia in their sample. This work shows that if these SNe~Ia are
standardized using
traditional light curve fitters such as SALT2, their distance estimates will have intrinsic biases
that could affect cosmology results. We are able to use the Twins Embedding to standardize SNe~Ia
using peak brightnesses and colors estimated from the SALT2 light curve fitter.
Targeted spectroscopic follow up campaigns could be used to localize SNe~Ia within the
Twins Embedding. This could potentially be done without the need for flux-calibrated
spectrophotometry, although further studies are necessary to determine how to handle host-galaxy
contamination. There is also additional information in
the light curve that is not captured by the SALT2 light curve model and that could potentially
be used to localize SNe~Ia in the Twins Embedding with more advanced light curve models.

Finally, using the Twins Embedding for standardization decreases correlations between distance
estimates and the properties of the SN~Ia host galaxies. We find that the step in host mass decreases from
\hoststepfitsaltmass~mag for SALT2 + $x_1$ standardization to \hoststepfitrbtlmass~mag for RBTL + Twins Embedding
standardization, and the step in LsSFR decreases from \hoststepfitsaltlssfr~mag to \hoststepfitrbtllssfr~mag.
Both of these decreases are significant at the $\hoststepdifflssfrsignificance\sigma$ level. These
results are for a sample where peculiar SNe~Ia have already been removed, so the decrease in host
step is not due to the SALT2 bias for 91T-like SNe~Ia. In fact, we find that including 91T-like SNe~Ia
in the sample decreases the size of the measured host step, since the 91T-like SNe~Ia in our sample have a host step
that is in the opposite direction of the one for the rest of the sample.
All of these results imply that future surveys need to measure properties of SNe~Ia beyond light curve
width and color if they are to produce robust cosmological measurements.

The code used to generate all of the results in this analysis is publicly available at
\url{https://doi.org/10.5281/zenodo.4670772}, and the data are available
on the SNfactory website at \url{https://snfactory.lbl.gov/snf/data/}.

\section{Acknowledgements}

We thank the technical staff of the University of Hawaii 2.2-m telescope, and Dan Birchall for observing assistance.
We recognize the significant cultural role of Mauna Kea within the indigenous Hawaiian community, and we appreciate
the opportunity to conduct observations from this revered site.
This work was supported in part by the Director, Office of Science, Office of High Energy Physics of the U.S.
Department of Energy under Contract No. DE-AC025CH11231.
Additional support was provided by NASA under the Astrophysics Data
1095 Analysis Program grant 15-ADAP15-0256 (PI: Aldering).
Support in France was provided by CNRS/IN2P3, CNRS/INSU, and PNC; LPNHE acknowledges support from LABEX ILP,
supported by French state funds managed by the ANR within the Investissements d’Avenir programme under
reference ANR-11-IDEX-0004-02.
Support in Germany was provided by DFG through
TRR33 “The Dark Universe” and by DLR through grants FKZ 50OR1503
and FKZ 50OR1602.
In China support was provided by Tsinghua University
985 grant and NSFC grant No. 11173017.
We thank the Gordon and Betty Moore Foundation for
their continuing support.
This project has received funding from the European Research Council (ERC)
under the European Union’s Horizon 2020 research and innovation programme
(grant agreement No. 759194 – USNAC).

\appendix

\section{Gaussian process regression} \label{sec:gaussianprocess}

In this analysis, we use Gaussian process (GP) regression to estimate the magnitude residuals of SNe~Ia as
a function of their position in the Twins Embedding. A stochastic process $P(x)$ is a
GP if for any finite set of points ${x_1, x_2, ..., x_n}$ the distribution
${P(x_1), P(x_2), ..., P(x_n)}$ is a multivariate normal distribution. A GP can be thought of as
a prior over a set of functions. By conditioning the GP on observations, we obtain a
posterior containing the set of functions that are consistent with the observations. For a detailed
discussion of GPs and their applications, see \citet{rasmussen06}.

A GP is uniquely defined by its mean function:
\begin{align}
    \mu(\vec{x}) = E[P(\vec{x})]
\end{align}
and its covariance function, or ``kernel'':
\begin{align}
    K(\vec{x}_1, \vec{x}_2) = E[(P(\vec{x}_1) - \mu(\vec{x}_1)) \times (P(\vec{x}_2) - \mu(\vec{x_2}))]
\end{align}

We denote this GP using the notation:
\begin{align}
    P(\vec{x}) \sim \mathcal{GP} \left(\mu(\vec{x}), K(\vec{x}, \vec{x})\right)
\end{align}

The choice of $\mu$ and $K$ determines how different functions are weighted in the prior of the GP.
There are several possible choices for the kernel. For our analyses, we use a Mat\'ern 3/2 kernel:
\begin{align}
    K_{3/2}(\vec{x}_1, \vec{x}_2; A, l) = A^2 \left(1 + \sqrt{3 \frac{\lVert \vec{x}_1 - \vec{x}_2 \rVert^2}{l^2}}\right) \exp\left(-\sqrt{3 \frac{\lVert \vec{x}_1 - \vec{x}_2 \rVert^2}{l^2}}\right)
\end{align}
The parameter $A$ describes the amplitude scale of the functions that will be produced
by the GP, and $l$ sets the length scale over which functions vary.
The Twins Embedding is three-dimensional, so each of the points $\vec{x}_i$ is also three dimensional.
By construction, Euclidean distances in the Twins Embedding directly map to the ``spectral distances''
between the original spectra. Hence we choose to use a single length scale for all of the dimensions,
and simply calculate the distances between any two points in the Twins Embedding using their
Euclidean distance, $\lVert \vec{x}_1 - \vec{x}_2\rVert$.

We chose to use a Mat\'ern 3/2 kernel as opposed to the RBF kernel that is commonly used
in the literature. The RBF kernel produces infinitely-differentiable functions which can lead to the
resulting models being unrealistically smooth \citep{stein99}. In contrast, the Mat\'ern 3/2 kernel
produces functions that are only once differentiable which is more realistic for some physical
processes. In practice, we find that our results are nearly identical with either kernel.

\bibliography{references}{}
\bibliographystyle{aasjournal}

\end{document}